\documentclass[11pt]{article}

\usepackage{fullpage, amsmath, amssymb, amsthm, bbm, color, dsfont}
\usepackage{graphicx,caption,subcaption,placeins}
\usepackage{setspace}
\usepackage{algorithm}
\usepackage[noend]{algpseudocode}
\usepackage[margin=1cm]{caption}
\usepackage[font=small]{caption}
\usepackage{booktabs,multirow}
\usepackage{float}

\makeatletter
\def\verbatim@font{\linespread{1}\normalfont\ttfamily}
\makeatother

\usepackage{enumerate}
\newtheorem{proposition}{Proposition}
\newtheorem{theorem*}{Theorem}
\newtheorem{proposition*}{Proposition}
\newtheorem{corollary*}{Corollary}

\newcommand{\bX}{{\bf X}}
\newcommand{\by}{{\bf Y}}
\newcommand{\errXY}{\textup{Err}_{\bX,\by}}
\newcommand{\errX}{\textup{Err}_\bX}
\newcommand{\errXt}{\textup{Err}_{\bX, \bX_{test}}}
\newcommand{\errin}{\textup{Err}_{\rm in}}
\newcommand{\E}{\mathbb{E}}
\newcommand{\var}{\textup{Var}}

\usepackage[bookmarks=true,bookmarksnumbered=true, hypertexnames=false,breaklinks=true]{hyperref}
\usepackage{authblk}

\title{Estimation of prediction error with known covariate shift}
\author{Hui Xu \thanks{Department of Statistics, Stanford University, huixu18@stanford.edu}, Robert Tibshirani\thanks{Department of Biomedical Data Science and Statistics, Stanford University, tibs@stanford.edu}}

\begin{document}
\maketitle

\abstract{In supervised learning, the estimation of  prediction error on unlabeled test data is an important task. Existing methods are usually built on the assumption that the training and test data are sampled from the same distribution, which is often violated in practice. As a result, traditional estimators like cross-validation (CV) will be biased and this may result in poor model selection. In this paper, we assume that we have a test dataset in which the feature values are available but not the outcome labels, and focus on a particular form of distributional shift called ``covariate shift''. We propose an alternative method based on parametric bootstrap of the target of conditional error $\errX$ \cite{bates2021cross}. Empirically our method outperforms CV for both simulation and real data example across different modeling tasks.}

\section{Introduction}
In predictive modeling, it is essential to estimate the generalization error on future test datasets. Given a particular model, such generalization error is implicitly dependent on the distribution from which the test data is drawn. Existing methods such as cross-validation (CV) usually rely on stationary assumptions between training and test data, which are often violated in practice due to time shift, location change, sampling bias, batch effects, etc. We consider estimation of generalization error when the covariate shift between training and test data is observed, and seek to improve upon existing methods by leveraging the additional covariate information of test data. 

More specifically, we focus on the scenario where covariates are observed for both training and test data, and the conditional distribution of outcome given covariates is the same for training and testing. This is known as ``covariate shift" \cite{sugiyama2007covariate,gretton2009covariate, tibshirani2019conformal}. In the case of time series data, the term ``virtual concept drift" \cite{gama2014survey, lu2018learning} is also used interchangeably. Our method is based on a slightly modified version of the target of inference $\errX$ \cite{bates2021cross}, which is the average prediction error of models fit on other unseen training datasets, and is shown to be an approximate estimand for CV when the stationary assumption between training and testing datasets is satisfied. We propose two ways to estimate the target of inference $\errX$ using either direct estimation or decomposition formula, resulting in two alternative estimators $\errX.dir$ and $\errX.dec$. 

Our method can be applied to a wide range of practical settings where covariate information of test data is readily available. For example, in genomics or proteomics settings, labeled data might be very expensive to obtain while unlabeled data is plentiful. Before committing to the cost of labeling more data, our method can be used to learn about the ability of the current model to generalize to unlabeled data. We will mainly discuss linear regression and classification, but our method can be applied to any predictive model.

% In predictive modeling, estimating predictive accuracy on unlabeled test points is an important task. Traditional methods to predict accuracy in supervised learning usually apply to situations where both covariate information and labels of future test points are unseen, and built on the assumption that training and test data come from the same population. However, generalizability from training to test data is rarely true in practice due to scarcity and limitations in labeled data used for training. On the other hand, we can usually observe covariates of test points on which prediction accuracy is of interest. Here we propose a method to estimate prediction accuracy on an unlabeled test data set with known covariate information, which could come from a different population from training data. 

\subsection{Illustration example}
\begin{figure}[!h]
\centering
\begin{minipage}{.42\textwidth}
  \centering
  \includegraphics[width=\linewidth]{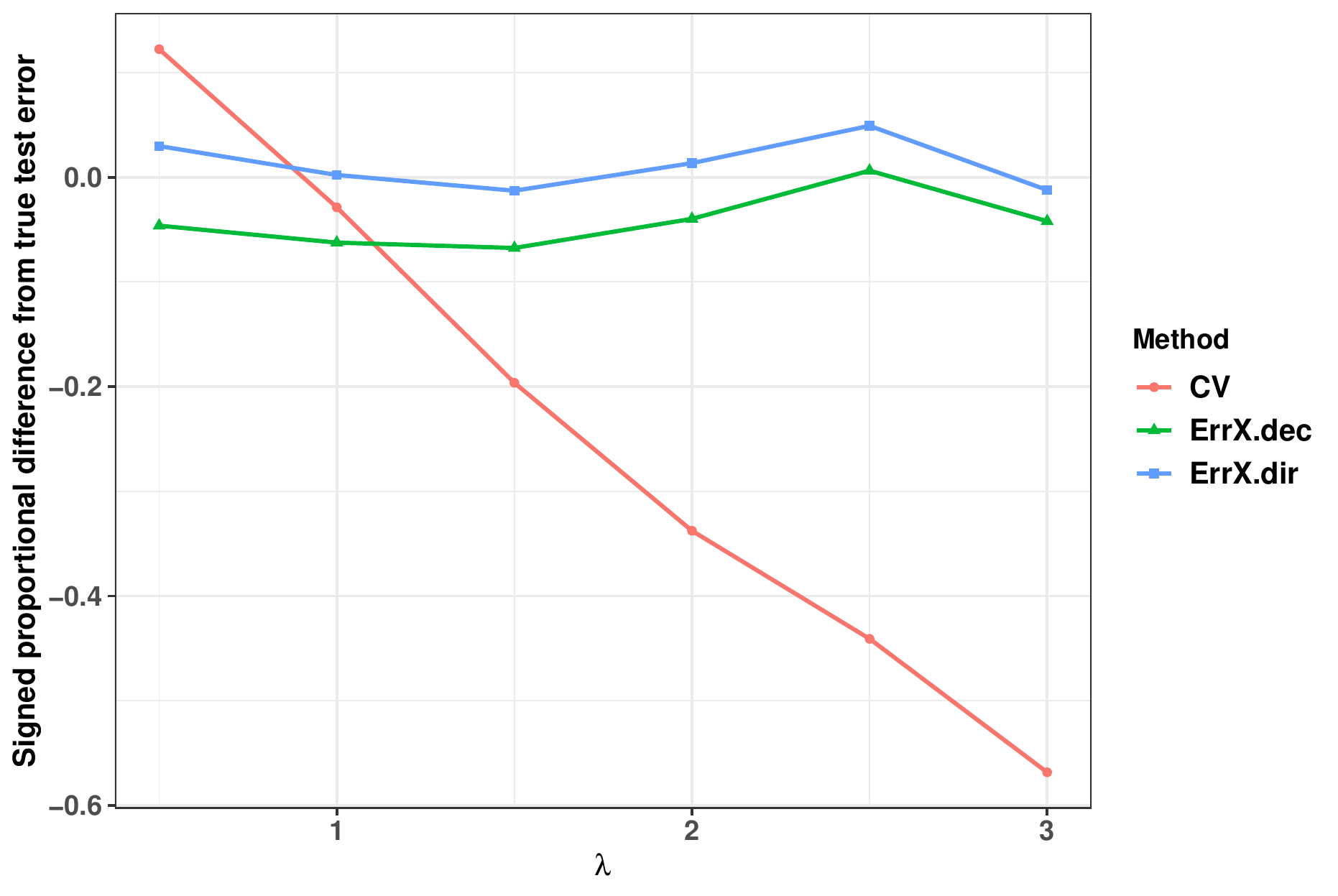}
%   \captionof{figure}{}
  \label{fig:test1}
\end{minipage}%
\begin{minipage}{.42\textwidth}
  \centering
  \includegraphics[width=\linewidth]{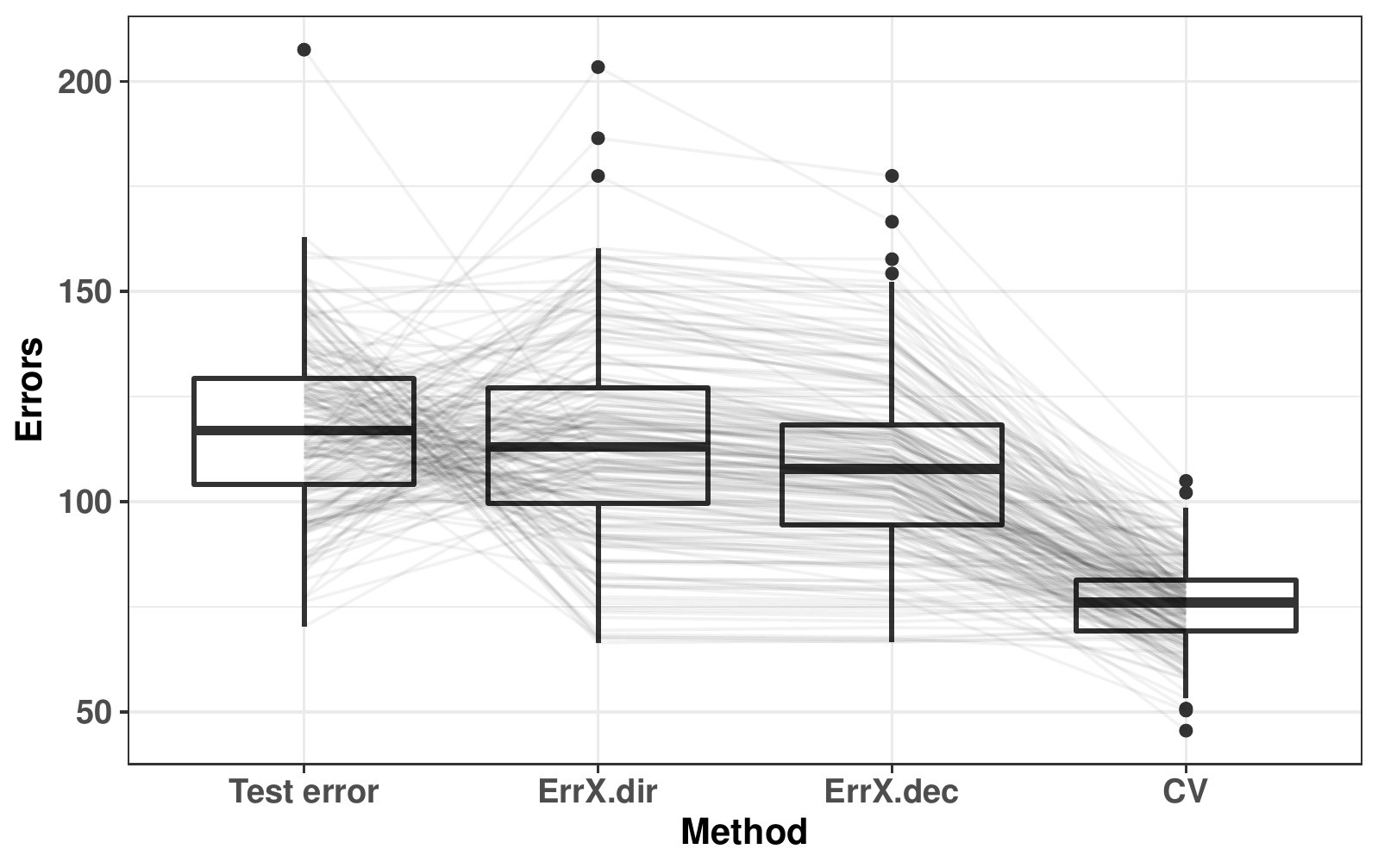}
%   \captionof{figure}{}
  \label{fig:test2}
\end{minipage}
\caption{\em Result of illustration example showing comparison between the two estimators that we propose (ErrX.dec, ErrX.dir) and cross validation (CV): The left-hand side is a plot of the average signed proportional difference between estimated prediction error and true test error under different magnitudes of covariate shifts according to parameter $\lambda$. The right-hand side plots details of the error distributions for $\lambda = 2$. The simulation results are averaged across $200$ simulations.}
\label{fig:intro}
\end{figure}

As a simple illustration, we compare our method with the widely-used technique of cross-validation (CV) in a simulated example. Consider a linear model $y_i= x_i^T\theta + \epsilon_i$, where $\epsilon_i$ are i.i.d   $\mathcal{N}(0,\sigma^2)$ and $x_i \in \mathbb{R}^p$ for $p = 50$ features. Suppose that we have a training data set of $100$ observations and an unlabeled test data set of $1000$ samples. We choose the training feature matrix $\bX$ to be comprised of independent and identically distributed (i.i.d.) standard normal variables, while the test feature matrix $\bX_{test}$ has entries of i.i.d $\mathcal{N}(0,\lambda^2)$ random variables, where $\lambda$ represents the amount of covariate shift from training data.  For each simulation setting, we choose $\lambda$ such that the signal-to-noise ratio (SNR) is approximately $3$. We fit the training data with Lasso using the
{\tt glmnet" package} \cite{glmnet} and evaluates its performance on the test set. 

In Figure \ref{fig:intro}, we compare the performance of error estimates in terms of the signed proportional difference from true test error, which can be seen as a measure of bias. Proportional difference is the ratio of the difference between error estimate and true test error divided by true test error. For example, the signed proportional difference for an estimator $\hat{e}$ and true test error $e$ is $(\hat{e} - e)/e$.
We can see that while CV predicts true test error well when the test data matrix is drawn from the same distribution as that of training data, its performance deteriorates significantly when there is covariate shift. On the other hand, the two estimators using our proposed method estimate true test error better across the spectrum of covariate shifts. 

\subsection{Related work}
% {\textbf{Methods for error prediction}}

There are two main categories for error prediction in a general framework without observing test data covariates, namely resampling techniques including cross-validation (\cite{efron1983estimating}) and related bootstrap-based techniques (\cite{efron1994introduction}, \cite{efron1997improvements}), as well as covariance penalty methods such as Mallow’s $C_p$ (\cite{doi:10.1080/00401706.1973.10489103}), AIC (\cite{akaike1974new}), BIC (\cite{schwarz1978estimating}), and Stein's unbiased risk estimate (\cite{stein1981estimation}) etc. In \cite{efron2004estimation}, the connection between the two theories was studied, and it was found that covariance penalties are a Rao-Blackwellized version of cross-validation. Cross-validation and related bootstrap techniques are nonparametric, while covariance penalties are model based, relying on the additive homoscedastic modeling assumption. 

The properties of CV are subtle and its estimand is elusive to precise definition. But it is generally agreed that just like covariance penalty methods, CV can also be considered as an estimate of expected prediction error, where the expectation is taken over both training data on future test points. There is plenty of work  (\cite{zhang1995assessing}, \cite{hastie2009elements}, \cite{yousef2020leisurely}, \cite{rosset2020fixed}, \cite{wager2020cross}, \cite{bates2021cross}) that discusses the estimand of CV.

% \noindent{\textbf{Target of $\errX$}}

In particular, instead of the instance-specific error for a particular training set, the estimand  $\errX$ was recently proposed by \cite{bates2021cross} as a better estimand for cross-validation error. $\errX$ is the average prediction error of models fit on other
unseen training outcomes drawn from the original superpopulation. The difference between CV and Mallow's $C_p$ is that while Mallow's $C_p$ targets in-sample error in the sense that future test points have the same covariates as training data, there is no such restriction for cross-validation.

% \noindent{\textbf{Generalization error with covariate information}}

Widely used methods for error prediction mentioned above are tailored to the classical setting where there is no distribution shift. And covariate information of test data is usually not used for error prediction. 
When the covariate values are  available for the target test data, a direct approach to estimate prediction on target is to use importance weighting (IW), which struggles when the supports of training and test data have little overlap especially in high dimensional settings (\cite{park2020calibrated}, \cite{stojanov2019low}). The authors in \cite{chen2021mandoline} proposed to mitigate the problem by guiding IW with prior information about directions in which distribution changes. When restricted to linear models, \cite{sugiyama2005generalization} derived an exact formula for prediction error on unlabeled test data, which can be estimated using parameter estimates from an importance weighted regression problem. Another method proposed recently \cite{yu2022predicting} to predict out-of-distribution error shows that there is high empirical correlation between projection norm and true test error. This method makes no assumptions on whether the conditional distribution shifts, but leverages on the feasibility of manually creating pseudo shifted datasets for calibration, which restricts its application to image or text classification.

% Distribution shift has long been identified as a challenge by the machine learning community. A large amount of work focus on developing model fitting and selection strategies that are robust or adapt to changes in data generating distribution (\cite{sugiyama2007covariate}, \cite{sagawa2019distributionally}, \cite{blanchet2019robust}, \cite{wen2014robust}, \cite{tibshirani2019conformal}, \cite{cauchois2020robust}, \cite{rothenhausler2021anchor}). However limited work is concerned with generalization error with distribution shift. 

% Target of inference
% {\color{red}{Stephen et al's paper estimating errx}}

\subsection{Organization of the paper}
The remainder of this paper is organized as follows. In Section 2, we clarify notations and describe our proposed method. In Section 3, we elaborate on our method in a few statistical models with corresponding simulation results. In Section 4, we demonstrate our method further with real-data example in crime rate prediction. In Section 5, we include some discussions of the advantages and limitations of our proposed method. 

\section{Description of method}
\label{method}
\subsection{Setting and notation}
We consider the supervised learning setting, where we have a training data set $\bX = (x_1,\ldots, x_n), \by = (y_1,\ldots, y_n)$ of $n$ observations drawn i.i.d from some joint distribution $P$. That is, $(x_i, y_i)_{i\in[n]} \overset{i.i.d}{\sim} P \in \mathcal{P} (\mathcal{X} \times \mathcal{Y})$, where $\mathcal{X} \subset \mathbb{R}^p$ and $\mathcal{Y} \subset \mathbb{R}$. Denote by $\delta_\bX$  the empirical distribution of covariates in $\bX$, $P_{X}$ as the marginal distribution of covariates, and $P_{Y|X}$ as the conditional distribution of $Y$ given $X$. Let $\hat{f}(x,\theta)$ be a function that predicts outcome $y$ from covariates $x \in \mathbb{R}^p$ using a parametric model with parameter $\theta \in \Theta \subset \mathbb{R}^d$. Let $\hat{\theta}$ be a function that maps values from $(\mathcal{X}\times \mathcal{Y})^n$ to parameter estimates in $\Theta$. Suppose that there is a new dataset consisting of i.i.d draws from $Q \in \mathcal{P}(\mathcal{X}\times \mathcal{Y})$, where $Q$ could be a different joint distribution but the conditional distribution of outcome given covariates remains the same. We are interested in predicting the test error, without access to the ground truth outcomes, as measured by a loss function, 
\begin{align*}
\ell: & \mathcal{Y} \times \mathcal{Y} \to \mathbb{R}_{\geq 0} \\
& (\hat{y},y) \mapsto \ell(\hat{y}, y)
\end{align*}
such that $\ell(y, y) = 0$ for all $y$. (For example, $\ell$ could be square error loss, misclassification error, or deviance.) 

The most intuitive target of inference for test error is the
\textit{out-of-sample error}, 
\begin{equation}
\label{errtest}
\errXY^{Q} := \E_{(x_0,y_0) \sim Q}\left[\ell(y_0, \hat{f}(x_0,\hat{\theta}(\bX,\by)) \mid \bX,\by\right],
\end{equation}
which is the expected loss when applying a model trained with $\bX,\by$ on a new data point $(x_0,y_0) \sim Q$. 
Suppose that we are given a particular unlabeled test set with $n_{test}$ samples and test covariates $\bX_{test}$. Similarly, let $\delta_{\bX_{test}}$ be the empirical distribution of covariates in $\bX_{test}$, and $Q_{Y|X}$ be the conditional distribution of outcome given covariates in the test population. Since we make the assumption that the conditional distribution remains unchanged and only deals with observed covariate shift, we have that $Q_{Y|X}= P_{Y|X}$. Notice that for an abuse of notation, we write $y \sim P_{Y|X=\bX}$ to mean the sampling of a random vector, $(y_1,\ldots, y_n) \sim P_{Y|X=x_1} \times \ldots \times P_{Y|X=x_n}$.

\subsection{Target}
We propose to estimate out-of-sample error in equation \eqref{errtest} by studying a similar averaged target first introduced in \cite{bates2021cross} as follows:  

\begin{align}
\begin{split}
\label{errX}
\errX^Q := \E[\errXY^Q|\bX] &= \E_{y \sim P_{Y|X= \bX}} \E_{(x_0,y_0) \sim Q}[\ell(y_0, \hat{f}(x_0,\hat{\theta}(\bX,y)) \mid \bX].
\end{split}
\end{align}

The motivation for using this target of inference is that when $P =Q$, the averaged target of $\errX^Q$ is closer to the true estimand of linearly invariant estimators (including cross-validation, data splitting, and Mallow's $C_p$) than the instance-specific $\errXY^Q$. Since $\errX^Q$ works well for test error prediction without distributional shift, we hope to generalize it to situations with observed covariate shifts. For more details, we refer interested readers to Theorem 1, Corollary 1 and 2 of \cite{bates2021cross}. 

Notice that since $Q$ is usually unknown, it is difficult to estimate $\errX^Q$ directly. But given test data covariates and the assumption of no conditional distribution shift, we can study a slightly modified version of the estimand, 
\begin{equation}
    \errXt^Q := \E_{y \sim P_{Y|X= \bX}} \E_{y_0 \sim P_{Y|X= x_0}}\E_{x_0 \sim \delta_{\bX_{test}}}[\ell(y_0, \hat{f}(x_0,\hat{\theta}(\bX,y)) \mid \bX].
    \label{errXt}
\end{equation}
The only difference between \eqref{errX} and \eqref{errXt} is that we replace the sampling from an unknown joint distribution $(x_0, y_0) \sim Q$ with its empirical counterpart. 

\subsubsection{Decomposition and connection to Mallow's $C_p$}

Since our new target $\errX^Q$ is a function of features in the training set, it has connections with \textit{in-sample error}, which is the target of estimation for traditional covariance-penalty based methods. Recall that \textit{in-sample error} $\errin(\bX)$  is the error for a fresh sample with the same covariates as training data. 
\begin{align}
\errin(\bX) &:= \mathbb{E}_{y, y' \sim P_{Y|X=\bX}}\left[\frac{1}{n} \sum_{i=1}^n \ell\left(y'_i, \hat{f}(x_i, \hat{\theta}(\bX,y))\right) \mid \bX\right] \notag \\
&= \E_{y \sim P_{Y|X= \bX}} \E_{y_0 \sim P_{Y|X= x_0}} \E_{x_0 \sim \delta_{\bX}} [\ell(y_0, \hat{f}(x_0,\hat{\theta}(\bX,y))\mid \bX].
\label{errin}
\end{align}

Notice that we can combine equations \eqref{errXt} and \eqref{errin} to obtain the following decomposition similarly as in \cite{bates2021cross}.
\begin{equation}
\label{decomp}
\begin{split}
\errXt^Q = \errin(\bX) +  \E_{y \sim P_{Y|X= \bX}} \E_{y_0 \sim P_{Y|X =x_0}} \left[\E_{x_0 \sim \delta_{\bX_{test}}}[\ell(y_0, \hat{f}(x_0,\hat{\theta}(\bX,y))] \right.  \\
 \left. - \E_{x_0 \sim \delta_\bX} [\ell(y_0, \hat{f}(x_0,\hat{\theta}(\bX,y))] \right].
\end{split}
\end{equation}

The value of this decomposition in \eqref{decomp} is that it offers an alternative way to estimate our target of $\errXt^Q$. For estimation of in-sample error $\errin$, we can use standard Mallows $C_p$ for linear models and bootstrap estimation for covariance penalty otherwise. For the remaining of the paper, we will denote estimators for the target in \eqref{errXt} as $\errX.dir$ (introduced later in Algorithm \ref{errx1}) and that for the decomposition in \eqref{decomp} as $\errX.dec$ (introduced later in Algorithm \ref{errx2}).  

% {\textcolor{red}{fill in details about the connection and previous literature, explain the motivation}}

\subsection{General methods of estimation}
This section considers estimation of targets introduced in the previous section under general supervised learning settings before elaborating in specific model applications later. We will illustrate separate procedures for direct estimation of target \eqref{errXt} and decomposition target \eqref{decomp} respectively. 

Our methods of estimation are based on the idea of parametric bootstrap. Let $P_{Y|X}^\theta$ be a parametric model and $\hat{\theta}$ be a parameter estimate. Then drawing parametric bootstrap samples $y \sim P_{Y|X}^{\hat{\theta}}$ means generating new outcomes for given covariate information based on the model parameterized by $\hat{\theta}$. For example, if $P^\theta_{Y|X}$ is a linear model parameterized by $\theta$ with Gaussian noise of mean $0$ and variance $\sigma^2$, then drawing parametric bootstrap sample $y \sim p^{\hat{\theta}}_{Y|X=x}$ means generating $y = x^T\hat{\theta} + \epsilon$ for $\epsilon \sim \mathcal{N}(0,\sigma^2)$ and parameter estimate $\hat{\theta}$. Again with abuse of notation, we write $Y \sim P^{\hat{\theta}}_{Y|X=\bX}$ to mean generating a random vector of independent parametric bootstrap samples $(y_1,\ldots, y_n) \sim P^{\hat{\theta}}_{Y|X=x_1} \times \ldots \times P^{\hat{\theta}}_{Y|X=x_n}$.

For direct estimation of the target in \eqref{errXt}, we illustrate the steps in Algorithm \ref{errx1}. After obtaining the initial parameter estimate from training data, we draw parametric bootstrap samples of new outcomes for both training and test covariates. For each bootstrap sample, we obtain an instance of the target by computing the loss between new test outcome and predicted outcome on test covariates using new training outcomes. The final estimate $\errX.dir$ can be obtained as an average of bootstrap errors.  

Similarly, for estimation of the decomposition target in \eqref{decomp}, we illustrate the steps in Algorithm \ref{errx2}. Given the same conditions and a suitable estimate for $\errin(\bX)$, algorithm \ref{errx2} estimates the difference term in \eqref{decomp} again by taking an average of bootstrap error differences. In addition to the steps in direct estimation, the only extra step is to compute the loss between new training outcomes and predicted outcomes for each bootstrap sample as bootstrap estimates of in-sample error. 

\begin{center}
\begin{minipage}{\linewidth}
\begin{algorithm}[H]
\caption{\em Direct estimation for $\errX.dir$}
\label{errx1}
\hspace*{\algorithmicindent} \textbf{Input}: training data $(\bX, \by)$, test covariates $\bX_{tests}$, loss $\l$, number of bootstrap samples $B$, fitting algorithm $\hat{\theta}(\cdot)$, parametric model $P^\theta_{Y|X}$
\begin{algorithmic}[1]
\State Fit a model on training data to obtain $\hat{\theta}(\bX, \by)$.
\For{each $b\in\{1,\ldots, B\}$}
\State Generate vectors of outcomes for training and test data,
\begin{align*}
    & \by^{(b)} \sim P^{\hat{\theta}(\bX,\by)}_{Y|X = \bX} \\
    & \by^{(b)}_{test} \sim P^{\hat{\theta}(\bX,\by)}_{Y|X = \bX^{test}}.
\end{align*}
 
\State Refit a model on bootstrap sample $\bX$, $\by^{(b)}$ to obtain $\hat{\theta}^{(b)} =\hat{\theta}(\bX, \by^{(b)})$. 
\State Compute 
\begin{equation*}
\label{dir}
\widehat{\errX}^{(b)} =\frac{1}{n_{test}}\sum_{i=1}^{n_{test}} l \left([\by^{(b)}_{test}]_i,\hat{f}(\bX^{test}_i,\hat{\theta}^{(b)}) ) \right).
\end{equation*}
\EndFor
\State Compute $\errX.dir = \frac{1}{B}\sum_{b=1}^B \widehat{\errX}^{(b)}$.
\end{algorithmic}
\hspace*{\algorithmicindent} \textbf{Output}: $\errX.dir$ 
\end{algorithm}
\end{minipage}
\end{center}

\begin{center}
\begin{minipage}{\linewidth}
\begin{algorithm}[H]
\caption{\em Estimate via decomposition $\errX.dec$}
\label{errx2}
\hspace*{\algorithmicindent} \textbf{Input}: training data $(\bX, \by)$, test covariates $\bX_{tests}$, loss $\l$, number of bootstrap samples $B$, fitting algorithm $\hat{\theta}(\cdot)$, parametric model $P^\theta_{Y|X}$
\begin{algorithmic}[1]
\State Fit a model on training data to obtain $\hat{\theta}(\bX, \by)$.
\State Compute estimate of $\errin(\bX)$ denoted as $\widehat{\errin}(\bX)$.
\For{each $b\in\{1,\ldots, B\}$}
\State Generate vectors of outcomes for training and test data,
\begin{align*}
    & \by^{(b)} \sim P^{\hat{\theta}(\bX,\by)}_{Y|X = \bX} \\
    & \by^{(b)}_{test} \sim P^{\hat{\theta}(\bX,\by)}_{Y|X = \bX^{test}}. 
\end{align*}
 
\State Refit a model on bootstrap sample $\bX$, $\by^{(b)}$ to obtain $\hat{\theta}^{(b)}= \hat{\theta}(\bX, \by^{(b)})$. 
\State Compute 
\begin{align*}
& \widehat{\errX}^{(b)} =\frac{1}{n_{test}}\sum_{i=1}^{n_{test}} l \left([\by^{(b)}_{test}]_i,\hat{f}(\bX^{test}_i,\hat{\theta}) \right)\\
& \widehat{\errin}^{(b)}(\bX)=\frac{1}{n}\sum_{i=1}^n l\left(\by_i^{(b)}, \hat{f}(\bX_i, \hat{\theta})\right). \label{dec}
\end{align*}

\EndFor
\State Compute $\errX.dec = \widehat{\errin}(\bX) +  \frac{1}{B}\sum_{b=1}^B \widehat{\errX}^{(b)} - \widehat{\errin}^{(b)}(\bX)$.
\end{algorithmic}
\hspace*{\algorithmicindent} \textbf{Output}: $\errX.dec$
\end{algorithm}
\end{minipage}
\end{center}

It remains to discuss possible ways to obtain suitable estimates of in-sample error $\widehat{\errin}(\bX)$. Notice that for ordinary least squares (OLS) with linear model, the well-known Mallows $C_p$ \cite{doi:10.1080/00401706.1973.10489103} is an unbiased estimate of in-sample error $\errin(\bX)$ by 
\begin{equation}
\label{cp}
\widehat{\mbox{Err}}^{(C_p)} := \frac{1}{n}\sum_{i=1}^n (y_i - \hat{f}(x_i, \hat{\theta}(\bX, \by)))^2 + \frac{2p\sigma^2}{n}. 
\end{equation}
For Lasso penalty in a linear model, we can replace the dimension of covariates $p$ with the number of nonzero coefficient estimates for estimating in-sample error via a degree of freedom argument. When dropping the linear model assumption, \cite{ye1998measuring} and \cite{efron2004estimation} give a more general form of covariance penalty identity for in-sample error, 
\begin{equation}
\label{cp_covariance}
\errin(\bX) = \E\left[\frac{1}{n}\sum_{i=1}^n \left(y_i - \hat{f}(x_i, \hat{\theta}(\bX, \by))\right)^2 \mid \bX\right] + \frac{2}{n} \sum_{i=1}^n \mbox{Cov} \left(y_i, \hat{f}(x_i, \hat{\theta}(\bX, \by))\mid \bX\right).
\end{equation}
This identity allows us to estimate in-sample error by parametric bootstrap. We refer interested readers to Appendix \ref{in-sample}. 

\section{Applications}
In this section, We illustrate the application of  the estimators $\errX.dir$ and $\errX.dec$ to specific settings, including  linear models (OLS and Lasso) and logistic regression for classification. We will discuss how the general methodology can guide us to estimate test error under various modeling assumptions and fitting algorithms with corresponding simulation results. 

\subsection{Linear model with Gaussian error}
First consider the setting of linear model with homoscedastic Gaussian errors,
$$
y_i = x_i^T \theta + \epsilon_i, \text{ where } \epsilon_i \stackrel{i.i.d}{\sim} \mathcal{N}(0,\sigma^2).
$$
Given a particular choice of loss function such as square error loss, the general method above can be directly applied. But there are subtleties involved depending on the initial fitting algorithm. For OLS there is a closed form solution for test error and our proposed estimators $\errX.dir$ and $\errX.dec$ are also unbiased. On the other hand, for model fitting with regularization such as the Lasso, while our estimators show advantage over traditional estimators such as CV in the presence of covariate shift, we will introduce additional debiasing modification that can help with our prediction. 

\subsubsection{Estimation for OLS}
If the initial fitting algorithm is OLS, then in addition to the estimates of $\errX.dir$ and $\errX.dec$ introduced above, we can also express the target in closed form similarly as in \cite{bates2021cross}. 

\begin{proposition}
\label{OLS_formula}
For linear models with the OLS fitting algorithm and squared error loss, assume in addition that the test data covariates are standardized so that $\mathbb{E}_{Q}[X] = 0$ and $\mbox{Var}_{Q}[X] = \Sigma$ of full rank. Then 
\begin{equation}
    \errX^Q = \sigma^2 + \frac{\sigma^2}{n} \mbox{tr}\left(\hat{\Sigma}^{-1} \Sigma\right) = \errin^P(\bX) + \frac{\sigma^2}{n} \left(\mbox{tr}\left(\hat{\Sigma}^{-1}\Sigma \right) - p\right),
\end{equation}
where $\hat{\Sigma} = \frac{1}{n} \sum_{i=1}^n x_i x_i^T$ is empirical covariance for training data.
\end{proposition}

For estimation using the general algorithm that we introduced in the above section, we can use unbiased estimators for $\sigma^2$ and $\errin(\bX)$, where $\hat{\sigma}^2 = \|\by- \bX \hat{\theta}\|_2^2 / (n-p)$ and $\widehat{\errin}(\bX) = \widehat{\mbox{Err}}^{(C_p)}$ in \eqref{cp}. 

\begin{proposition}
\label{OLS-unbiased}
For linear models with the OLS fitting algorithm and squared error loss, assume in addition that the test data covariates are standardized so that $\mathbb{E}_{Q}[X] = 0$ and $\mbox{Var}_{Q}[X] = \Sigma$ of full rank. Then the estimators $\errX.dir$ and $\errX.dec$ in the algorithms \ref{errx1} and \ref{errx2} are unbiased for the target $\errX^Q$.
\end{proposition}

\begin{figure}[!h]
\centering
\begin{minipage}{.45\textwidth}
  \centering
  \includegraphics[width=\linewidth]{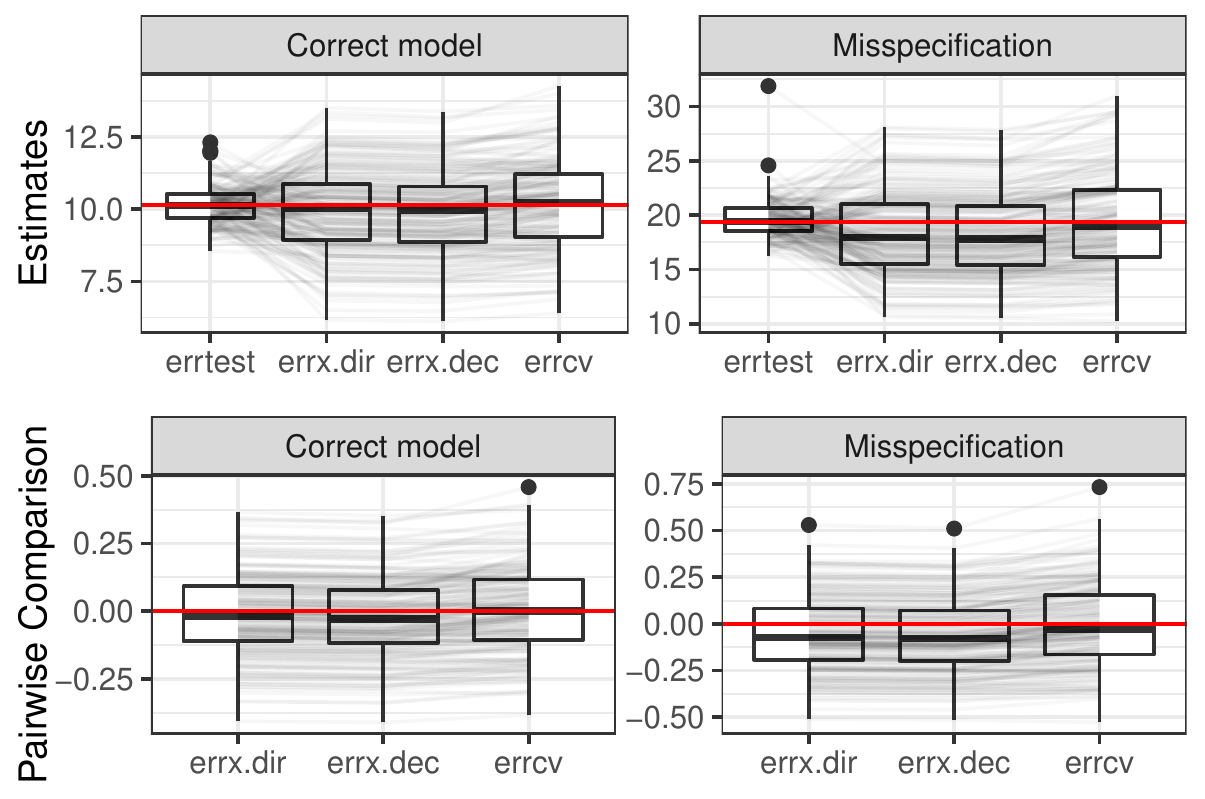}
%   \captionof{figure}{}
  \label{fig1:OLS}
  \subcaption{No covariate shift}
\end{minipage}%
\begin{minipage}{.45\textwidth}
  \centering
  \includegraphics[width=\linewidth]{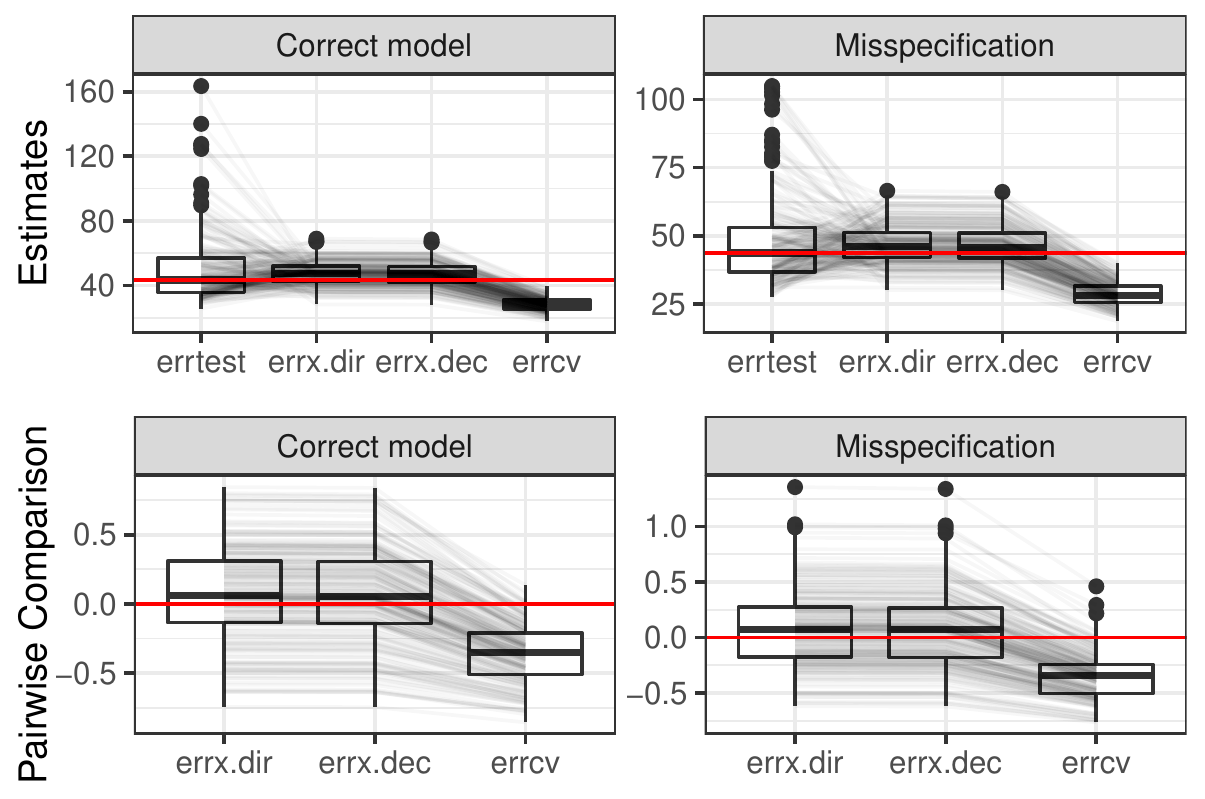}
%   \captionof{figure}{}
  \label{fig2:OLS}
  \subcaption{Covariate shift}
\end{minipage}
\caption{\em Estimates of prediction error for OLS under the absence (left) and presence (right) of covariate shift. For estimates, comparisons are among true test error, $\errX$ estimates ($\errX.dir$ and $\errX.dec$), and cross validation. For each setting, a pairwise comparison is included in the second row corresponding to the proportion of deviation from true test error for each of the three estimates.}
\label{fig:OLS}
\end{figure}

For simulation, we consider the setting with $n=100$ observations of $p=10$ features for training, $n_{test}=1000$ observations of unlabeled test data, and coefficient vector of $4$ nonzero entries with equal strength of $2$. The training feature matrix consists of i.i.d entries drawn from $\mathcal{N}(0,1)$. We consider two situations, one without covariate shift and one with covariate shift. The detailed setup is as follows. 

\begin{enumerate}
    \item No covariate shift: The feature matrix of test data are comprised of i.i.d entries drawn from $\mathcal{N}(0,1)$. We choose $\sigma = 3$ so that signal-to-noise ratio (snr) is approximately $2$. For sensitivity to model misspecification, we include a quadratic transformation to $1/3$ of the feature coordinates.
    \item Covariate shift: The feature matrix of test data are comprised of i.i.d entries drawn from $\mathcal{N}(2,2)$. We choose $\sigma = 5$ so that signal-to-noise ratio (snr) is approximately $2$. For sensitivity to model misspecification, we include a quadratic transformation to $1/5$ of the feature coordinates.
\end{enumerate}

Notice that we choose different transformations such that the change in test error with or without taking the transformation into account is approximately $50 - 100\%$. From Figure \ref{fig:OLS} it can be seen that while both cross validation and $\errX$ estimation predicts true test error well when there is no covariate shift (left plot), our method of $\errX$ estimation performs better than CV in the presence of covariate shift (right plot). Under model misspecification, while CV demonstrates slightly more robustness without covariate shift, the bias in CV error estimation in the presence of covariate shift outweighs the robustness advantage. 

% Assume without loss of generality that for test data covariates, $\mathbb{E}(X) = 0$ and $\mbox{Var}(X) = \Sigma$. Then equations \eqref{errx_s} and \eqref{de2} can be further simplified as the following. 

% \begin{align}
% \label{OLS_errx}
% %\sigma^2 \left(1 + \mbox{tr}\left(\Sigma \left(\bX^T\bX \right)^{-1} \right)\right),
% \sigma^2 + \frac{\sigma^2}{n} \mbox{tr}\left(\hat{\Sigma}^{-1} \Sigma\right)
% \end{align} 
% \begin{align}
% \label{OLS_decomp}
% \errin(\bX) + \frac{\sigma^2}{n} \left(\mbox{tr}\left(\hat{\Sigma}^{-1}\Sigma \right) - p\right),
% \end{align}
% respectively, where $\hat{\Sigma} = \frac{1}{n} \sum_{i=1}^n X_i X_i^T$ is empirical covariance for training data. \textcolor{magenta}{(HX) Todo: Need to add reference to Stephen et al's paper}.

\subsubsection{Estimation for Lasso}
There is no explicit formula for the error target if we use fitting algorithms with regularization, such as the Lasso. However, our general method in algorithms \ref{errx1} and \ref{errx2} can be used. By \cite{reid2016study}, we know that a good estimate for $\sigma^2$ in Lasso regression is 
\begin{equation*}
\hat{\sigma}^2 = \frac{1}{n-\hat{s}_{\hat{\lambda}}}\|\by- \bX \hat{\theta}_{\hat{\lambda}}\|_2^2,
\end{equation*}
where $\hat{\theta}_{\hat{\lambda}}$ is the Lasso estimate at regularization parameter $\hat{\lambda}$ selected via cross-validation, and $\hat{s}_{\hat{\lambda}}$ is the number of nonzero elements in $\hat{\theta}_{\hat{\lambda}}$. For estimation of $\errin(\bX)$, we can use covariance penalty identity in equation \eqref{cp_covariance}. Notice that in Lasso regression \cite{tibshirani2012degrees},
$$
 df = \frac{1}{\sigma^2}\sum_{i=1}^n \mbox{Cov} \left(y_i, \hat{f}(x_i,\hat{\theta}(\bX, \by))\right) = \hat{s}_{\hat{\lambda}}.
$$
Therefore we can use the following unbiased estimator for in-sample error
\begin{equation}
    \widehat{\errin}(\bX) = \frac{1}{n}\sum_{i=1}^n (y_i - \hat{f}(x_i, \hat{\theta}(\bX, \by)))^2 + \frac{2 \hat{s}_{\hat{\lambda}}\hat{\sigma}^2}{n}.
\end{equation}

An additional subtlety for Lasso fitting algorithm is that in addition to the general method listed in Algorithms \ref{errx1} and \ref{errx2}, we need bias correction steps in order to achieve better prediction accuracy. Since the Lasso estimator is biased, the parametric bootstrap step will carry on the bias, requiring corrections to the output estimators $\errX.dir$ and $\errX.dec$. Here we propose two existing methods of bias correction for predicting test error of Lasso as an example, i.e. (1) Multiplicative bootstrap correction and (2) Relaxed Lasso. The details are as follows. 
\begin{enumerate}
\item \textbf{Multiplicative bootstrap bias correction: }Multiply estimators $\errX.dir$ and $\errX.dec$ by a constant shrinking factor $c$. Let $\hat{\theta}(\bX,\by)$ and $\hat{\theta}(\bX,\by^{(b)})$ denote the fitted parameters from initial model and bootstrap samples. Then we propose to choose, 
$$
c = \frac{\|\hat{\theta}(\bX,\by)\|^2}{\frac{1}{B}\sum_{b=1}^B \|\hat{\theta}(\bX,\by^{(b)})\|^2}.
$$
The intuition is that we adjust for the scaling factor between the true parameter $\theta$ and learned parameter $\hat{\theta}$ by that between $\hat{\theta}$ and refitted parameter after bootstrap. 

\item \textbf{Relaxed Lasso correction: }Use relaxed Lasso fit on the initial training data to form parametric bootstrap samples. The idea is that we want to reduce the bias between true parameter $\theta$ and that used in generating parametric bootstrap samples. 
\end{enumerate}

\begin{figure}[!h]
\centering
\begin{minipage}{.45\textwidth}
  \centering
  \includegraphics[width=\linewidth]{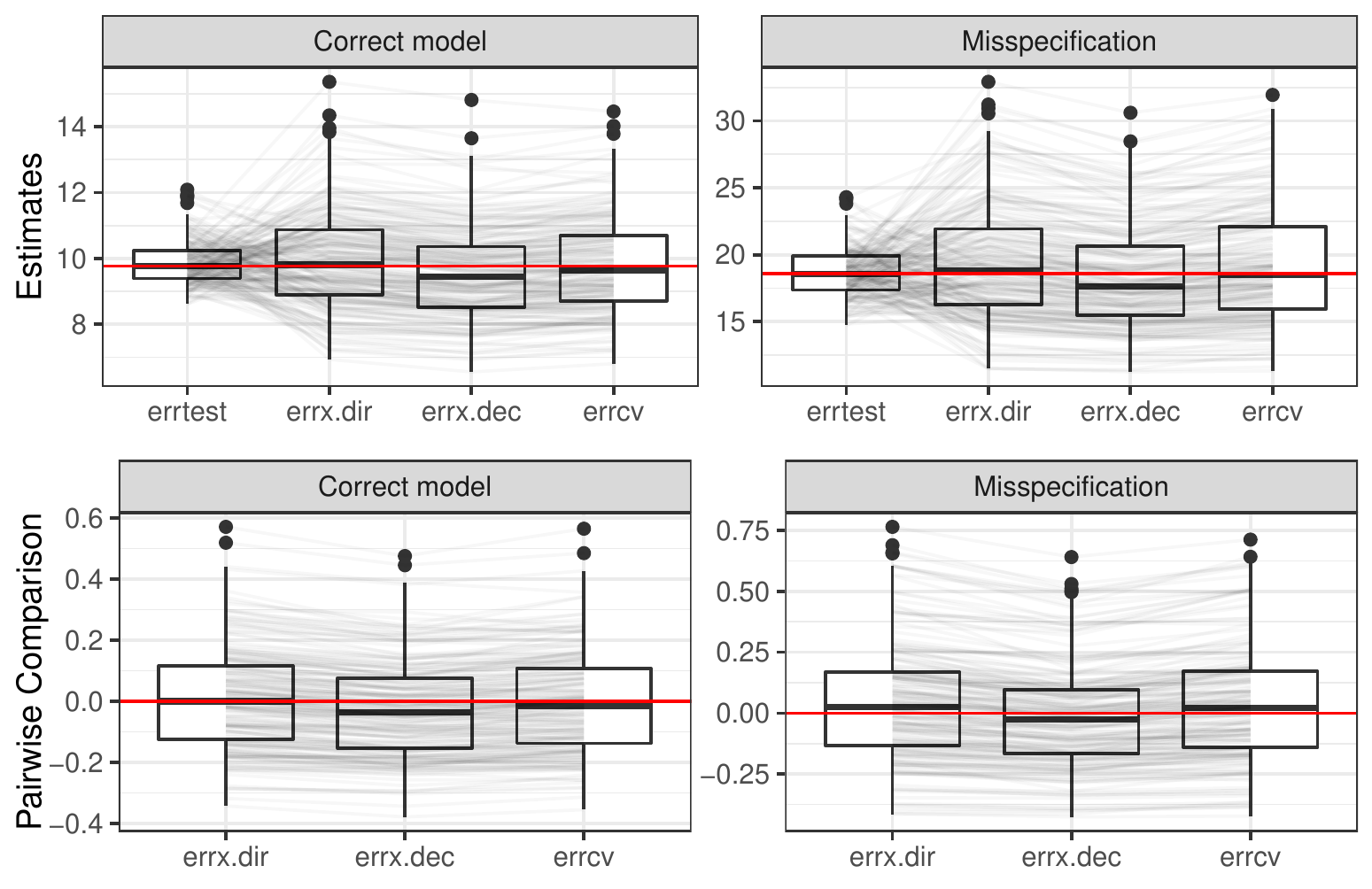}
%   \captionof{figure}{}
  \label{fig1:glmnet_lowdim}
  \subcaption{No covariate shift}
\end{minipage}%
\begin{minipage}{.45\textwidth}
  \centering
  \includegraphics[width=\linewidth]{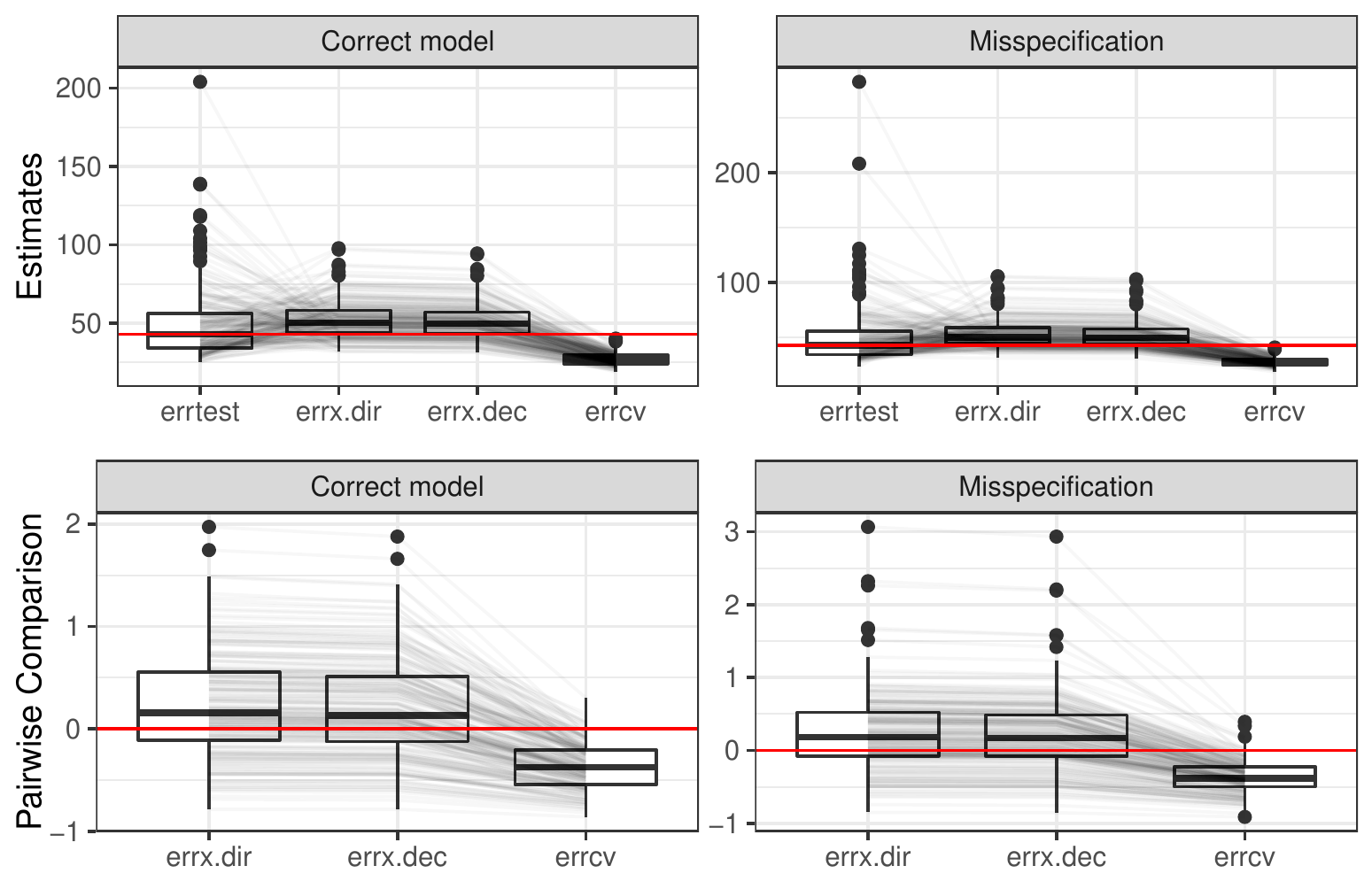}
%   \captionof{figure}{}
  \label{fig2:glmnet_lowdim}
  \subcaption{Covariate shift}
\end{minipage}
\caption{\em Estimates of errors for linear regression with Lasso penalty in the lower dimensional setting ($p = 10$) under the absence(left) and presence(right) of covariate shift. For estimates, comparisons are among true test error, $\errX$ estimates ($\errX.dir$ and $\errX.dec$), and cross validation. For each setting, a pairwise comparison is included in the second row corresponding to the proportion of deviation from true test error for each of the three estimates.}
\label{fig:glmnet_lowdim}
\end{figure}

For simulation of error estimation in Lasso fitting algorithm using the {\tt glmnet} R package \cite{glmnet} with multiplicative bias correction, we consider two scenarios: a low dimensional setting as in OLS $p = 10$ as a benchmark, and a higher dimensional setting $p = 50$. The simulation results for lower dimensional setting are presented in Figure \ref{fig:glmnet_lowdim}. 

\begin{figure}[!h]
\centering
\begin{minipage}{.45\textwidth}
  \centering
  \includegraphics[width=\linewidth]{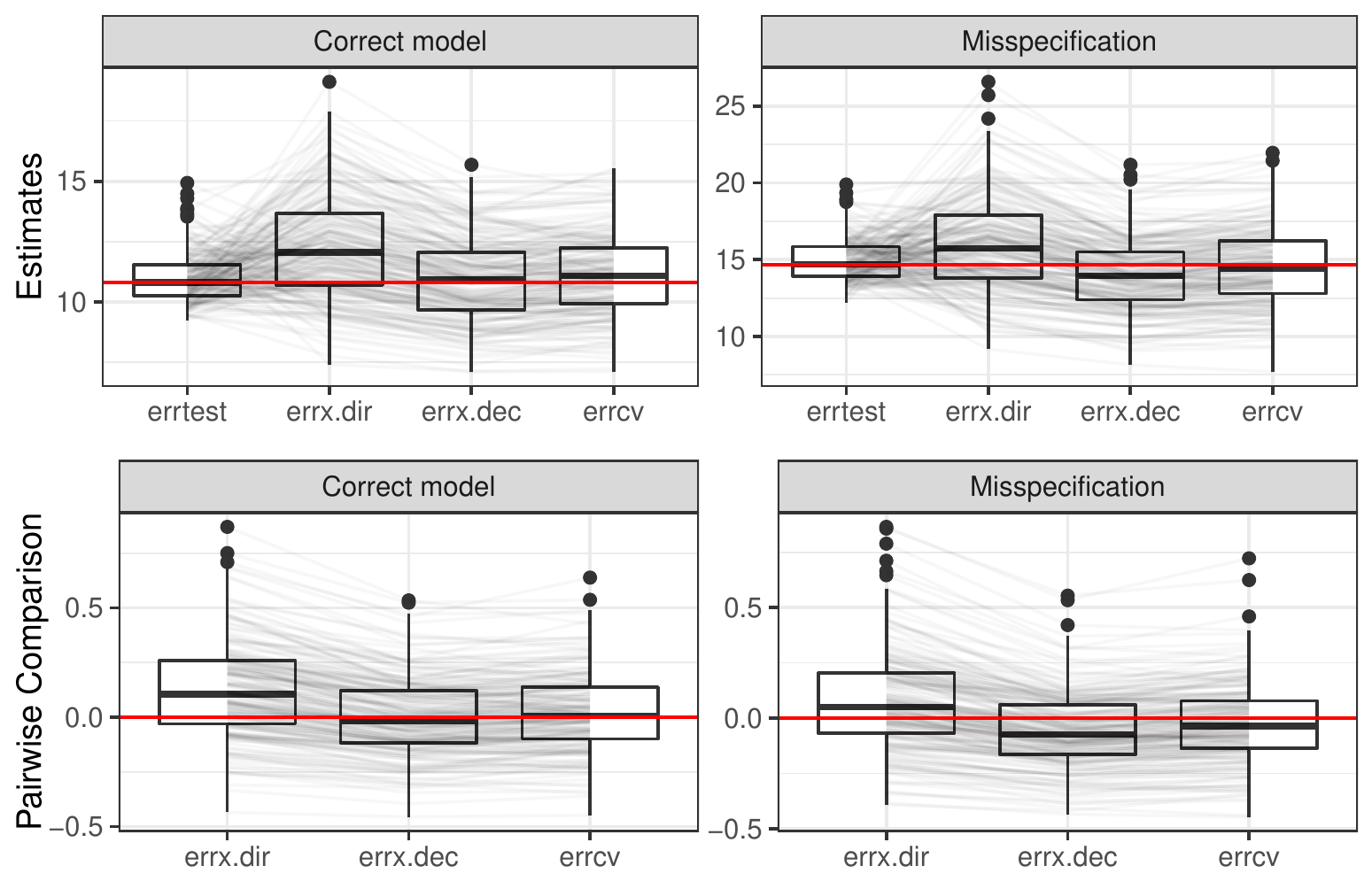}
%   \captionof{figure}{}
  \label{fig1:OLS}
  \subcaption{No covariate shift}
\end{minipage}%
\begin{minipage}{.45\textwidth}
  \centering
  \includegraphics[width=\linewidth]{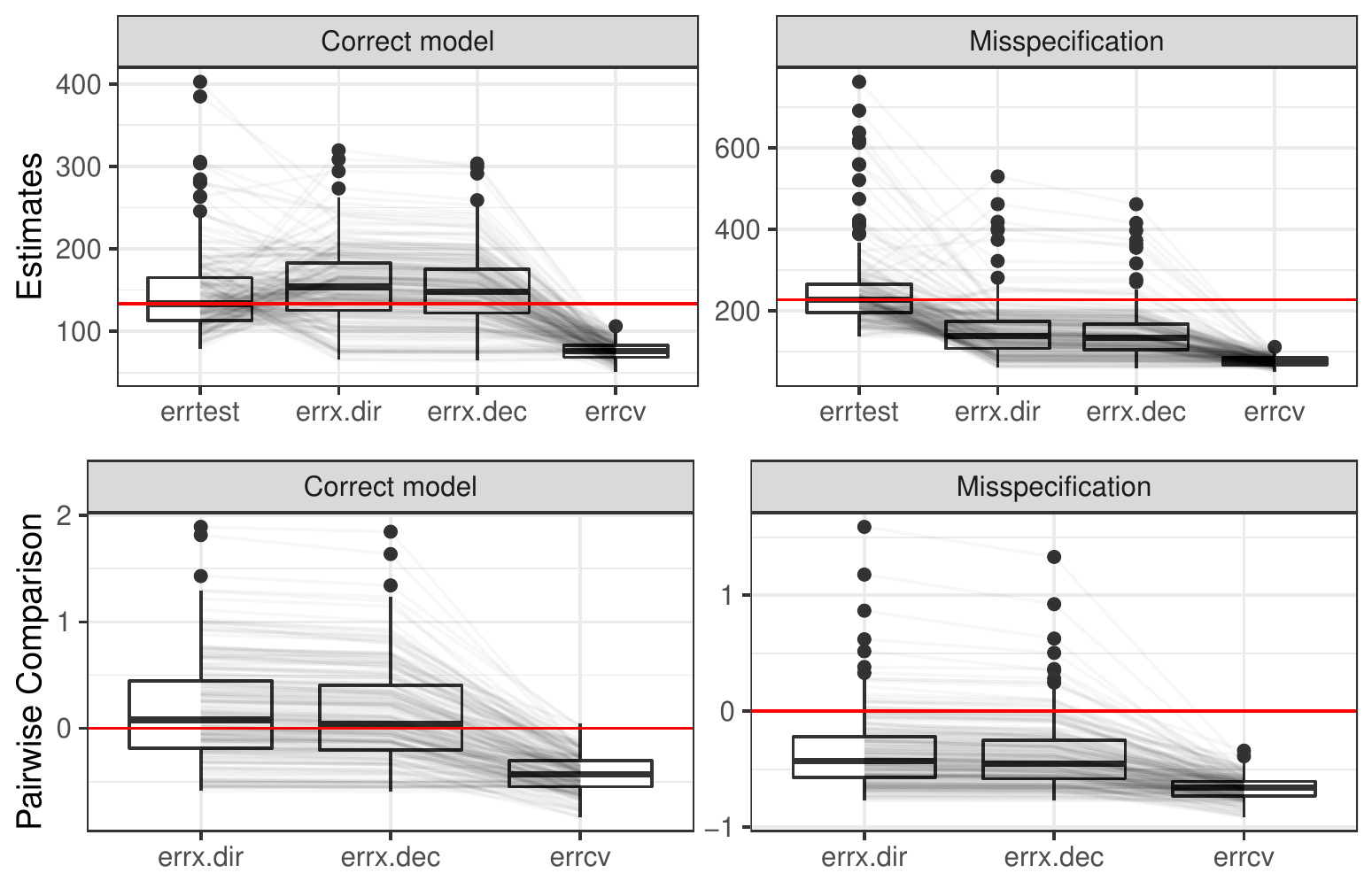}
%   \captionof{figure}{}
  \label{fig2:OLS}
  \subcaption{Covariate shift}
\end{minipage}
\caption{\em Estimates of errors for linear regression with Lasso penalty in the higher dimensional setting ($p = 50$) under the absence(left) and presence(right) of covariate shift. For estimates, comparisons are among true test error, $\errX$ estimates ($\errX.dir$ and $\errX.dec$), and cross validation. For each setting, a pairwise comparison is included in the second row corresponding to the proportion of deviation from true test error for each of the three estimates.}
\label{fig:glmnet}
\end{figure}

We also consider the higher dimensional setting with $n = 100$ observations of $p = 50$ features for training, $n_{test} = 1000$ observations of unlabeled test data, and coefficient vector of 5 nonzero entries with equal strength of $2$. Again we draw training feature matrix i.i.d from $\mathcal{N}(0,1)$ and consider the following two situations.
\begin{enumerate}
    \item No covariate shift: The feature matrix of test data are comprised of i.i.d entries from $\mathcal{N}(0,1)$. We choose $\sigma = 3$ so that $snr \approx 2$. For model misspecification, we include a quadratic transformation to the first $1/5$ feature coordinates. 
    \item Covariate shift: The feature matrix of test data are comprised of i.i.d entries drawn from $\mathcal{N}(2,2)$. We choose $\sigma = 8 $ so that $snr \approx 2$. For model misspecification, we include a transformation to the first $1/10$ feature coordinates by taking them to the $1.5$ power so that the amount of model misspecification is around $30\%$. 
\end{enumerate}
The simulation results for the higher dimensional setting are presented in Figure \ref{fig:glmnet}. 

It can be seen that for both settings of $p=10$ and $p=50$, the results of error prediction are analogous to those in OLS setting. Both error estimates $\errX.dir$ and $\errX.dec$ predicts true test error well and perform better than CV in the presence of covariate shift. 

\subsection{Generalized Linear Model (GLM)}
Our method of estimating $\errX$ can similarly be applied to other nonlinear generalized linear models (GLM). For Bernoulli observations as an example, we can replace square error loss with a suitable loss for binary classification such as counting error or binomial deviance, and use logistic regression as the fitting algorithm. It is worth noting that as part of the procedure to produce the estimator $\errX.dec$, we need to estimate in-sample error, which can be obtained with general covariance penalties \cite{efron2004estimation}. We provide additional details of in-sample error estimation in Appendix \ref{in-sample}. 

For simulation of $\errX$ estimation in nonlinear GLM, we consider a sparse logistic model
$$
P(Y_i =1 \vert X_i = x_i)= \frac{1}{1 + e^{-x_i^T\theta}},
$$
with $n = 200$ observations and two cases for the number of features: a low dimensional setting $p = 10$, and a higher dimensional setting $p = 50$. The training feature matrix consists of i.i.d entries drawn from $\mathcal{N}(0,1)$. We are interested in the comparison of different error estimates using counting error, both with and without covariate shift. For covariate shift, we draw i.i.d test data from $\mathcal{N}(3,1)$ and subsample training data so that the training labels are imbalanced with ratio of $3$. We chose the sparsity and signal strength so that signal-to-noise ratio is approximately $3$. 

Similarly as in the case of linear regression, we need to apply bias correction to the Lasso parameter estimates. Here we use relaxed Lasso correction for bootstrap in both $\errX$ estimation and in estimating in-sample error.

\begin{figure}[!h]
\centering
\begin{minipage}{.45\textwidth}
  \centering
  \includegraphics[width=\linewidth]{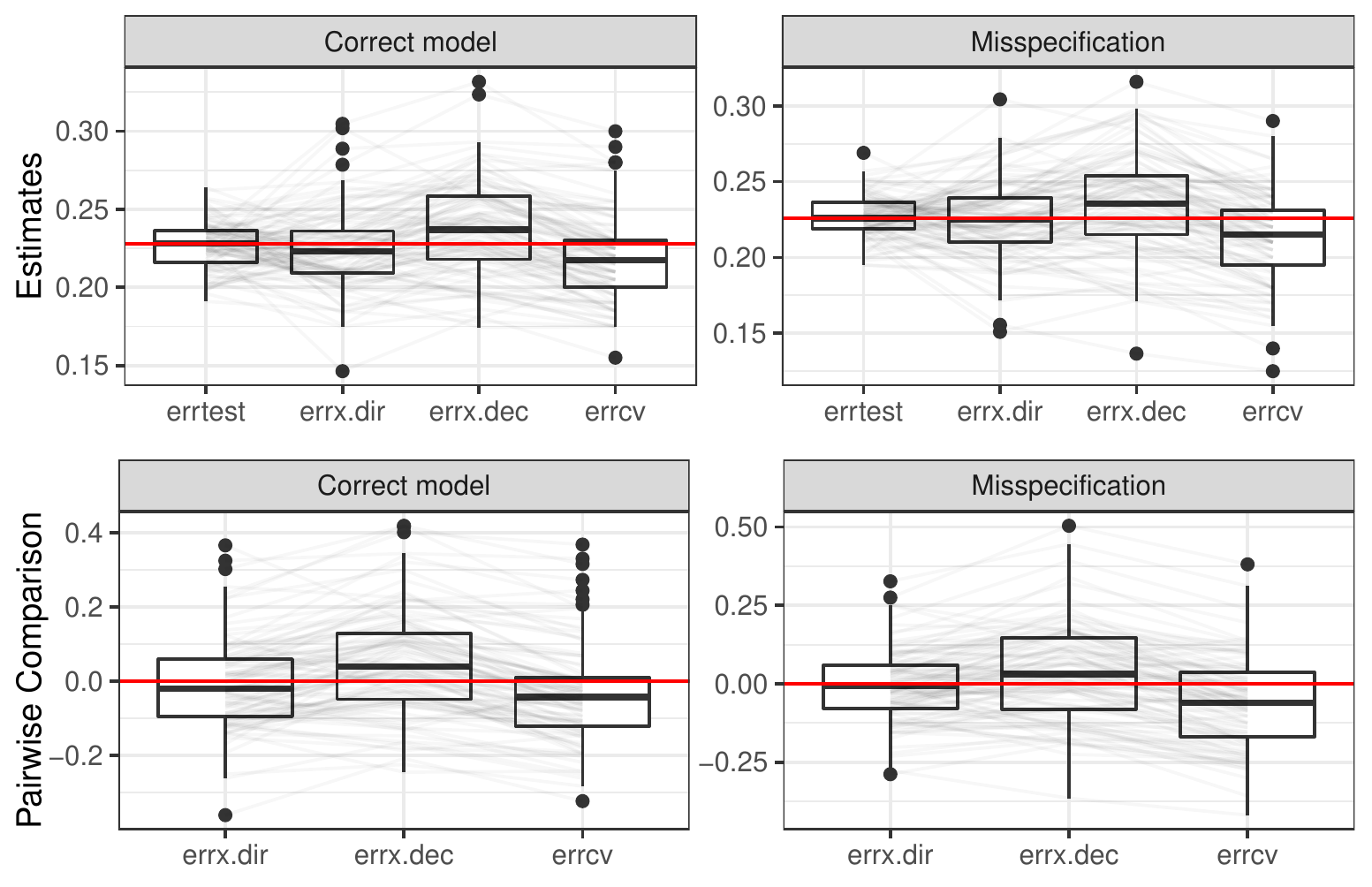}
%   \captionof{figure}{}
  \label{fig1:logistic_lowdim}
  \subcaption{No covariate shift}
\end{minipage}%
\begin{minipage}{.45\textwidth}
  \centering
  \includegraphics[width=\linewidth]{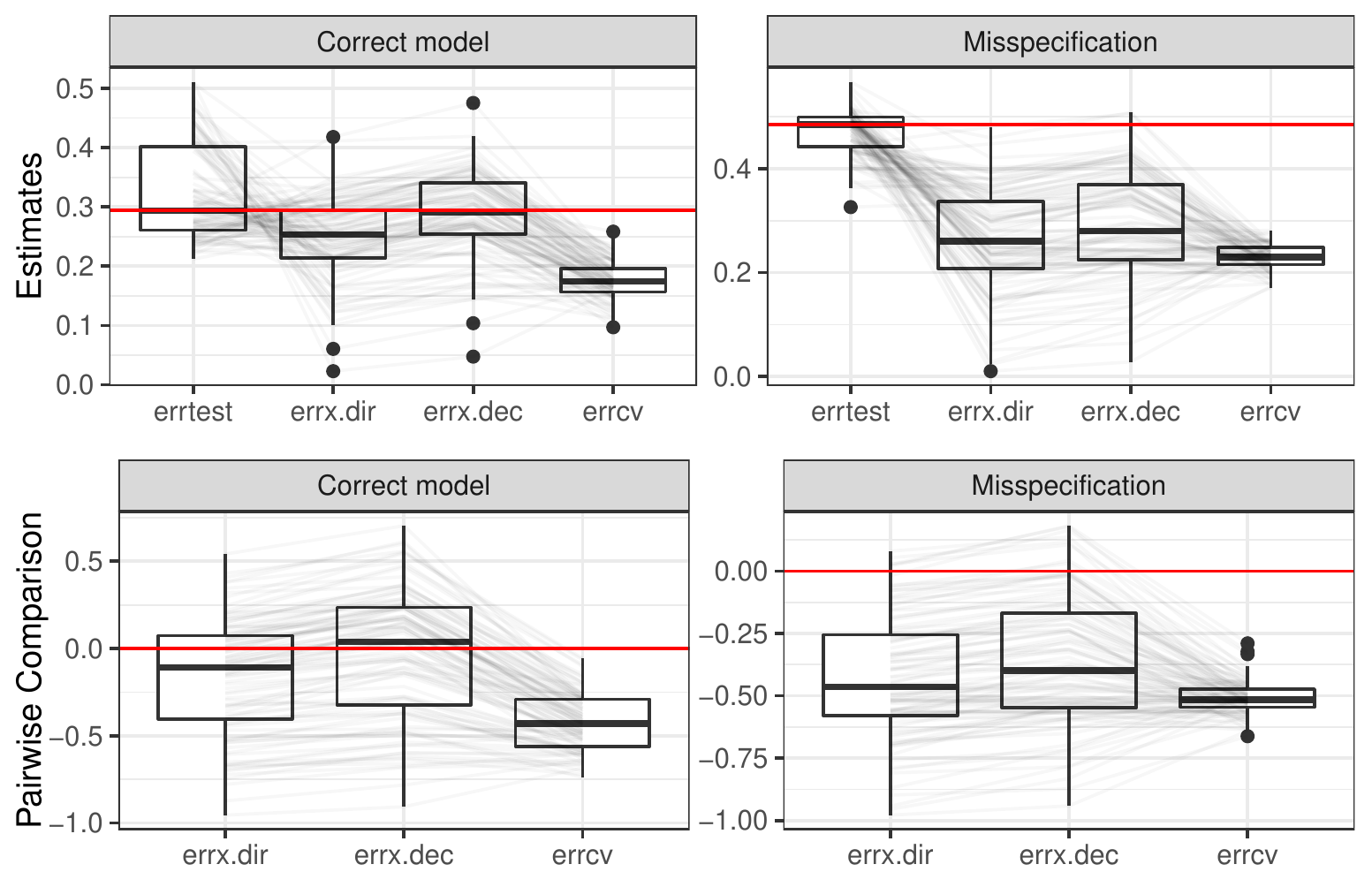}
%   \captionof{figure}{}
  \label{fig2:logistic_lowdim}
  \subcaption{Covariate shift}
\end{minipage}
\caption{\em Estimates of errors for logistic regression with Lasso penalty in the lower dimensional setting ($p = 10$) under the absence(left) and presence(right) of covariate shift. For estimates, comparisons are among true test error, $\errX$ estimates ($\errX.dir$ and $\errX.dec$), and cross validation. For each setting, a pairwise comparison is included in the second row corresponding to the proportion of deviation from true test error for each of the three estimates.}
\label{fig:logistic_lowdim}
\end{figure}

\begin{figure}[!h]
\centering
\begin{minipage}{.45\textwidth}
  \centering
  \includegraphics[width=\linewidth]{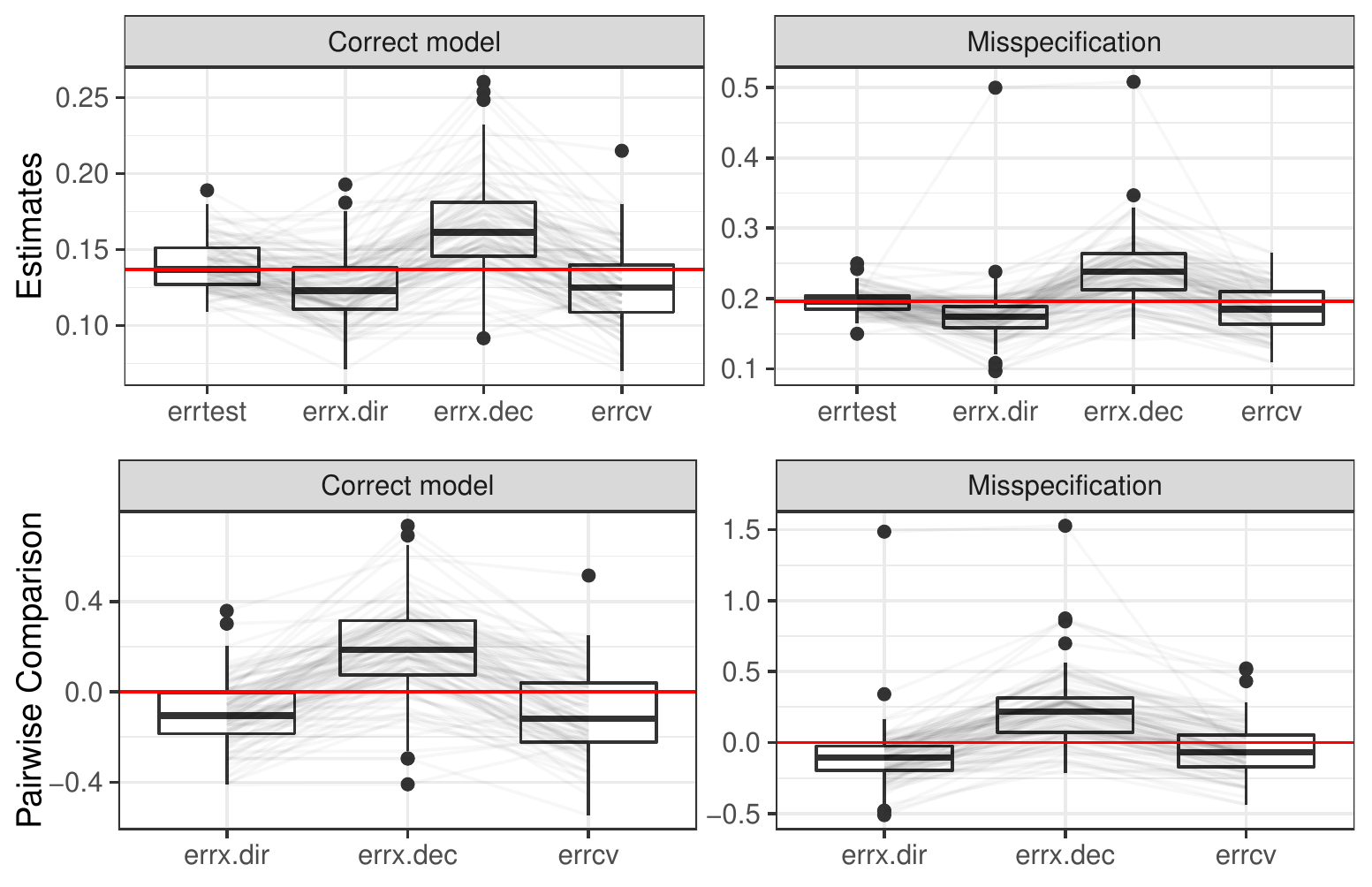}
%   \captionof{figure}{}
  \label{fig1:logistic_highdim}
  \subcaption{No covariate shift}
\end{minipage}%
\begin{minipage}{.45\textwidth}
  \centering
  \includegraphics[width=\linewidth]{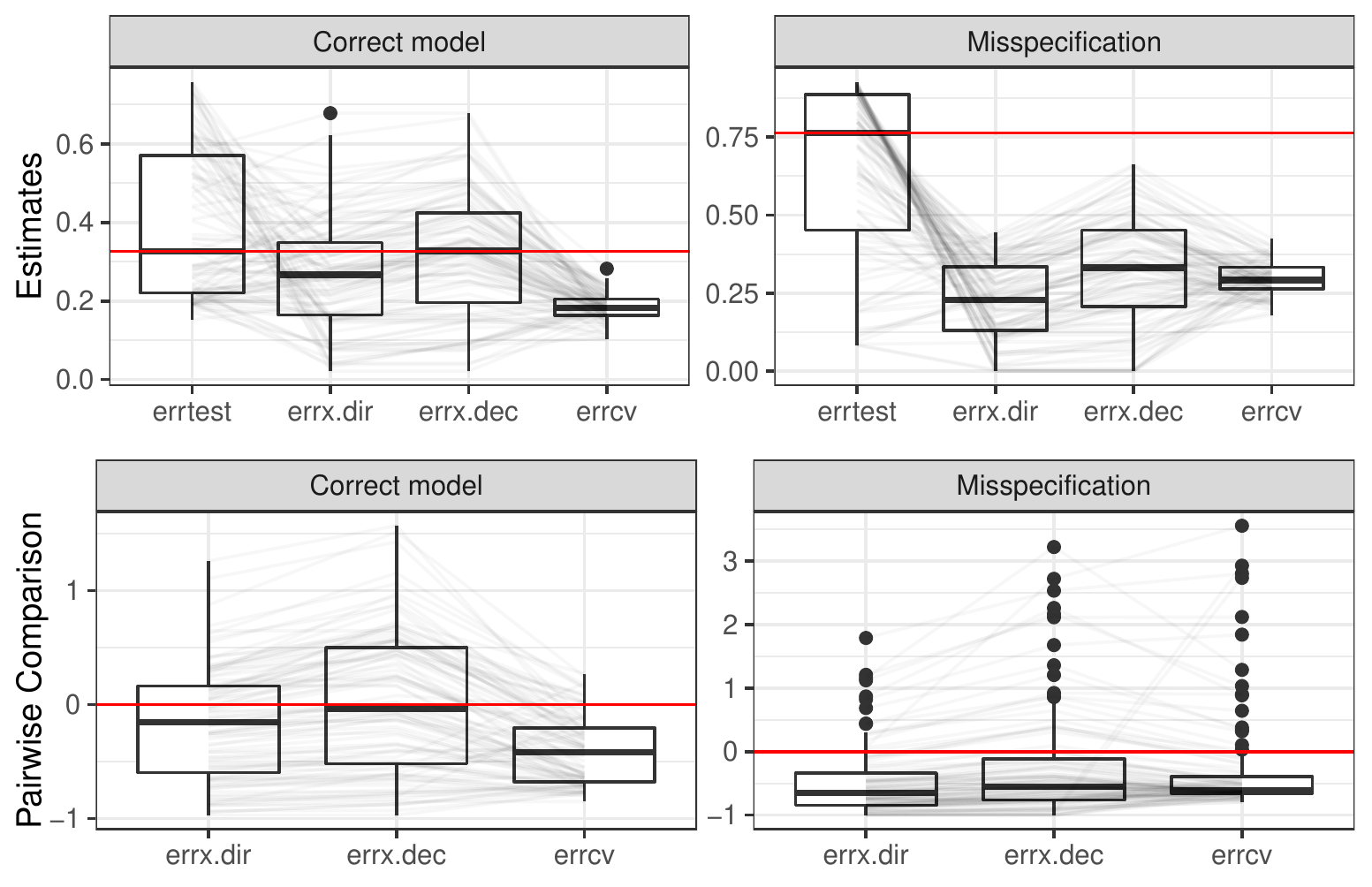}
%   \captionof{figure}{}
  \label{fig2:logistic_highdim}
  \subcaption{Covariate shift}
\end{minipage}
\caption{\em Estimates of errors for logistic regression with Lasso penalty in the higher dimensional setting ($p = 50$) under the absence(left) and presence(right) of covariate shift. For estimates, comparisons are among true test error, $\errX$ estimates ($\errX.dir$ and $\errX.dec$), and cross validation. For each setting, a pairwise comparison is included in the second row corresponding to the proportion of deviation from true test error for each of the three estimates.}
\label{fig:logistic_highdim}
\end{figure}

The simulation results for $p = 10$ and $p = 50$ are given in Figure \ref{fig:logistic_lowdim} and Figure \ref{fig:logistic_highdim}, respectively. Similarly as in OLS and linear regression with Lasso penalty, the two proposed estimates $\errX.dir$ and $\errX.dec$ recovers true test error better than CV in the presence of covariate shift. It has to be acknowledged that when there is model misspecification along covariate shift, none of the error estimates resembles true test error. However, the model misspecification that we introduced is an artificial and drastic one, which makes about $1/5$ of the covariates in the linear model quadratic instead. Good performance under model misspecification is not a reasonable expectation, and the setting is included only as a caution for application under model misspecification. 

\subsection{Summary of simulation}
\begin{table}[!h]
\centering
\begin{tabular}{lccccc} 
\toprule
& \multirow{2}{*}{OLS}  & \multicolumn{2}{c}{Linear (Lasso)} & \multicolumn{2}{c}{Logistic (Lasso)} \\ \cmidrule(lr){3-4}\cmidrule(lr){5-6}
& & $p=10$  & $p=50$ & $p=10$   & $p=50$   \\\midrule
CV  & 0.766 & 0.452 & 0.481 & 0.453 & 0.541   \\ \midrule
$\errX.dir$  & 0.0645 & -0.058 & -0.124 & 0.232 & 0.367\\ \midrule
$\errX.dec$  & 0.0655 & -0.0371 & -0.0569 & 0.109 & 0.225  \\ \bottomrule
\end{tabular}
 \caption{\em Comparison of average signed difference between error estimates and actual test error for above simulation settings with covariate shift. The values in the table are standardized by the mean test error. Smaller absolute values are better. Multiplicative correction is used for linear regression with Lasso penalty, and relaxed Lasso correction is used for logistic regression.}
 \label{table-simulation-shift}
\end{table}
The summary of result comparisons under covariate shift can be found in Table \ref{table-simulation-shift}. Since the metric used is the average signed difference between error estimates and true test error, the table compares estimators based on bias. We see a similar trend in table entries as in detailed box plot comparisons above. Under covariate shift, our two proposed estimators of $\errX.dir$ and $\errX.dec$ perform much better than CV across all simulation settings. We also include a summary table of comparison in terms of other metrics that take variance into account in Appendix \ref{table-simulation-others}.

\section{Real data example}
\subsection{States crime rate}
We analyze a public data set on yearly state crime rates from 1977 to 2014, obtained initially from John Donahue of Stanford Law School. The data set contains $42$ demographic variables as predictors for the outcome of violent crime rate, with a total number of $1887$ entries. We first split the data set into two parts for training and testing. We fit a linear model with Lasso penalty on the training set and apply different methods for error estimations, including our proposed estimators $\errX.dir$ and $\errX.dec$, as well as cross-validation. We then compare different error estimates with true test error evaluated on the test set outcomes. We consider three different scenarios for splitting as follows and summarize the mean squared difference between error estimates and true test error in Table \ref{table-crime}.  

\begin{enumerate}
    \item \textbf{Random half splits}: Data is randomly assigned to training and testing, regardless of the state and year. 
    \item \textbf{Random half splits by states}: States are randomly assigned to training and testing. Data entries belonging to the same state are kept in the same fold in cross-validation. 
    \item \textbf{Two-means clustering by states}: Two-means clustering is applied on centroid of all states to split states into training and testing. Data entries belonging to the same state are kept in the same fold in cross-validation. 
\end{enumerate}

\begin{table}[!h]
\centering
\begin{tabular}{lccccc} 
\toprule
& \multirow{2}{*}{CV}  & \multicolumn{2}{c}{$\errX.dir$} & \multicolumn{2}{c}{$\errX.dec$} \\ \cmidrule(lr){3-4}\cmidrule(lr){5-6}
& & Multi  & Relax & Multi    & Relax   \\\midrule
Random half splits    & 5.93e-3 & 6.42e-3 & 6.11e-3 & 5.93e-3 & 5.96e-3   \\ \midrule
Random half splits by states  & 0.572 & 0.438 & 0.468 & 0.415 & 0.473 \\ \midrule
Two-means clustering by states  & 0.904 & 0.561 & 0.708 & 0.561 & 0.710  \\ \bottomrule
\end{tabular}
 \caption{\em Comparison of mean squared difference between error estimates and actual test error for various splitting settings. Each mean squared difference is averaged over 200 splits. Both $\errX.dir$ and $\errX.dec$ are calculated via two bias correction methods, i.e. multiplicative correction and relaxed Lasso correction. Smaller values are better. Note that in multiplicative correction, we cap the error above zero and restrict the multiplicative factor from being too large.}
 \label{table-crime}
\end{table}

\begin{figure}[!h]
\centering
\includegraphics[width=0.8\linewidth]{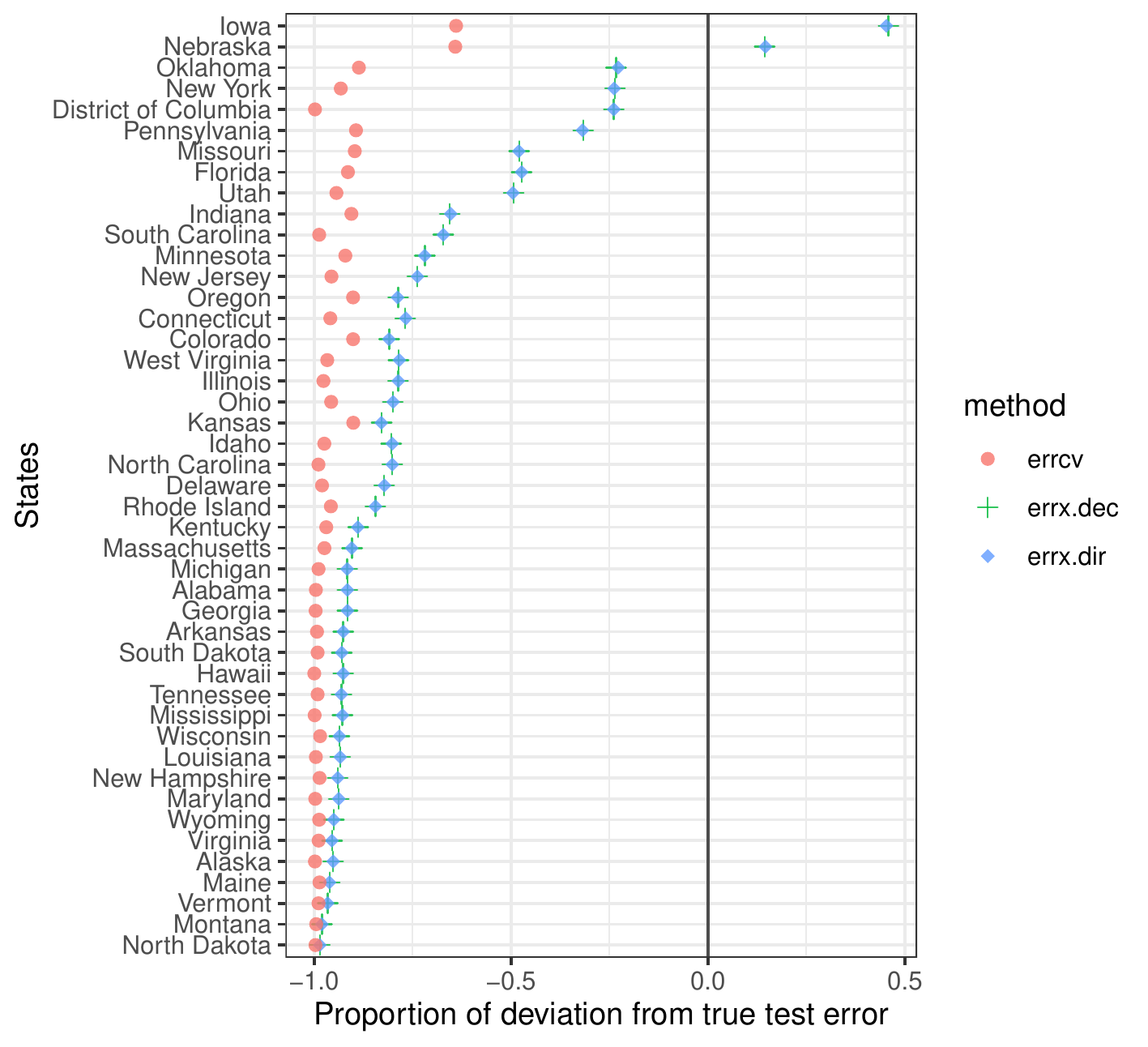}
\caption{\em  Comparison of proportional difference between error estimates and actual test error for each state as test set. The fixed training set include California, Washington, Nevada, New Mexico, Arizona, and Texas. Multiplicative bias correction method is used to estimate $\errX.dir$ and $\errX.dec$.}
\label{plot-crime}
\end{figure}

In the first case of random half splits, we expect no distribution shift, where all error estimates considered are close to the actual test error. In the second case of random half splits by states, we expect some covariate shift as well as possible distribution shift due to different relations between demographic predictors and outcome across different states. It can be seen that $\errX.dir$ and $\errX.dec$ perform slightly better than cross-validation. The recovery of true test error is not perfect due to potential violation of the assumption of no conditional distribution shift. In the third case of two-means clustering by states, we try to maximize covariate shift between training and testing. Estimators $\errX.dir$ and $\errX.dec$ perform much better than cross-validation despite potential conditional distribution shift. 

We also analyze the crime rate data by fixing a training set consisting of data from a few states in the west including California, Washington, Nevada, New Mexico, Arizona, and Texas. We then compare different error estimates and actual test error by traversing over the remaining test states in Figure \ref{plot-crime}. The estimates of $\errX.dir$ and $\errX.dec$ perform strictly better than cross-validation in every test case, especially in Nebraska, Iowa, Oklahoma, New York, District of Columbia, Pennsylvania, Missouri, Florida, and Utah, possibly due to smaller shift in conditional distribution of crime rate given predictor variables. 

\subsection{Image classification}
We apply ErrX method on the K-class image classification task CIFAR10 \cite{krizhevsky2009learning} and compare with two other existing methods: Average confidence (ConfScore) (\cite{hendrycks2016baseline}), and Projection Norm (ProjNorm) (\cite{yu2022predicting}). The method of ErrX as described in Algorithm \ref{errx1} and \ref{errx2} are broadly applicable beyond prediction with (regularized) linear regression or generalized linear models. Here we consider wrapping ErrX with neural network for image classification in computer vision. 

For test data involving distribution shift, we consider both the orginal version and some adapted version of the common corruptions dataset (\cite{hendrycks2019benchmarking}). Since labels for images remain the same after corruption, they may not satisfy the exact covariate shift assumption. We provided an adapted common corruptions dataset via relabeling to ensure only covariate shift in test data, generated by the following procedure.  

\begin{enumerate}
    \item Split CIFAR-10 training data randomly into two parts, labeling set $\{(x_i, y_i)\}_{i \in \mathcal{I}_1}$ and training set $\{(x_i, y_i)\}_{i \in \mathcal{I}_2}$, where $\mathcal{I}_1 \cup \mathcal{I}_2 = \{1,\ldots, n\}$ are non-overlapping set of indices. 
    \item Starting from ResNet18 architecture, fine tune the network using labeling set $\{(x_i, y_i)\}_{i \in \mathcal{I}_1}$ to obtain base model $1$ with parameter $\hat{\theta}_{1}$.
    \item Replace labels in the training set $\{(x_i, y_i)\}_{i \in \mathcal{I}_2}$ and OOD CIFAR-10 corruption set $\{\tilde{x}_i, \tilde{y}_i\}_{1:m}$ with an estimated (pseudo) label using base model $1$, i.e. 
    \begin{align*}
        & y_{i} \sim \textup{Multinomial}(f_1(x_{i}, \hat{\theta}_1), \ldots, f_K(x_{i}, \hat{\theta}_1)), i \in \mathcal{I}_2\\
        & \tilde{y}_i \sim \textup{Multinomial}(f_1(\tilde{x}_i, \hat{\theta}_1),\ldots, f_K(\tilde{x}_i, \hat{\theta}_1)), i = 1,\ldots, m
    \end{align*}
    \item Starting from ResNet18 architecture, fine tune the network using pseudo-labeled set $\{(x_i, y_i)\}_{i \in \mathcal{I}_2}$ to obtain base model $2$ with parameter $\hat{\theta}_2$.
    \item Predict the performance of base model $2$ on pseudo-labeled OOD test set $\{\tilde{x}_i\}_{1:m}$ with ErrX/ProjNorm/Confscore method. 
\end{enumerate}

For initial training, we sample $50000$ data points from CIFAR10 dataset and train for $20$ epochs with ResNet18 architecture \cite{he2016deep} pre-traineded on ImageNet \cite{deng2009imagenet}. For fine-tuning, we use SGD with warm restart with learning rate $10^{-3}$, momentum $0.9$, minibatch size of $128$, and cosine learning rate decay \cite{loshchilov2016sgdr}. For refitting in computing ProjNorm and ErrX estimators, we use the same optimizer and fine-tuning with $500$ iterations. For OOD (out-of-distribution) testing data, we sample $10000$ test samples from the common corruptions dataset with maximum severity level $5$. We chose the maximum corruption severity level in order to have larger separation between training distribution and test distribution. Our code is adapted from open source code in \cite{yu2022predicting}. 

\begin{figure}[H]
\begin{minipage}{0.28\textwidth}
         \centering
    \includegraphics[width=\textwidth]{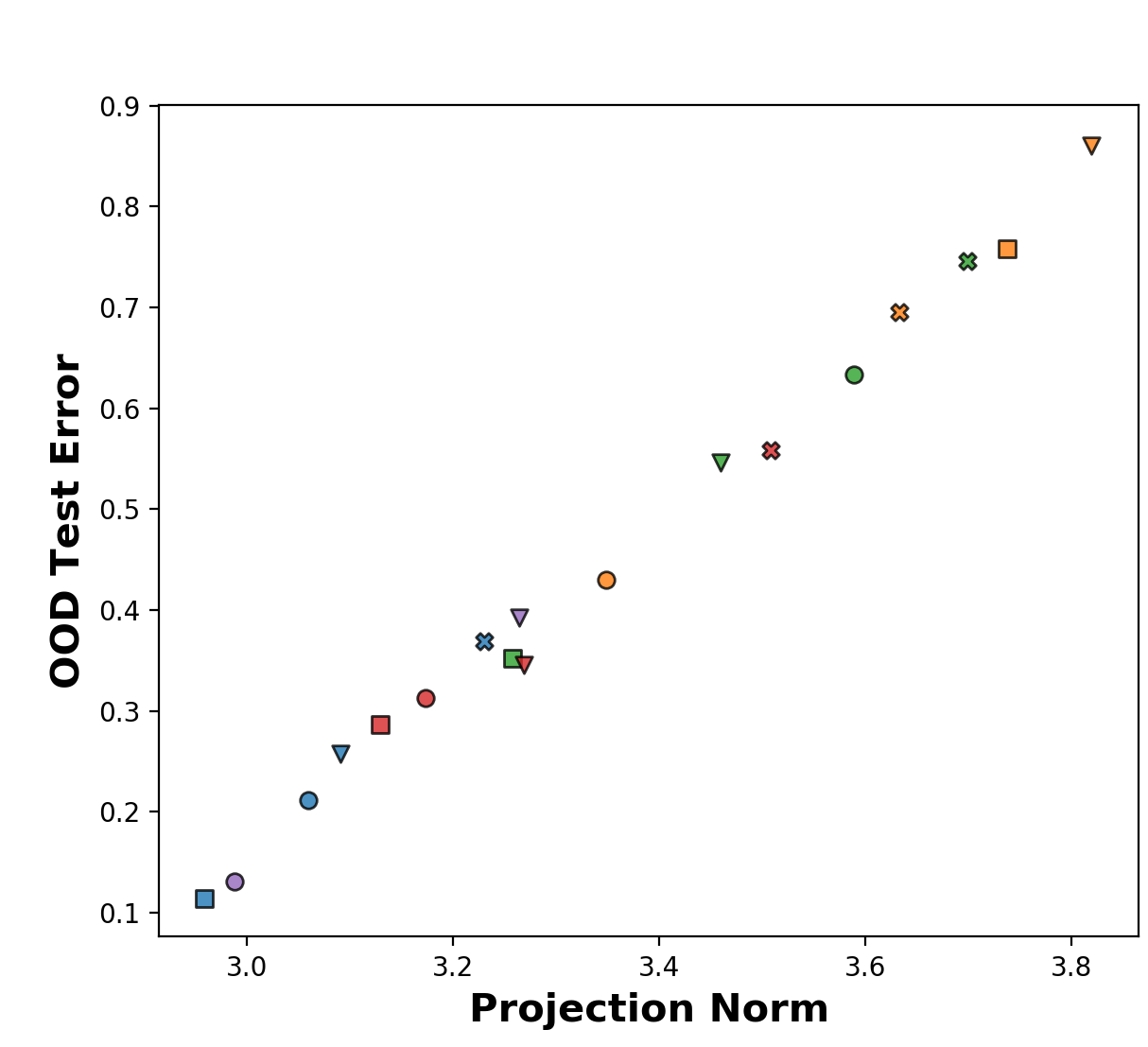}
    \end{minipage}
        \begin{minipage}{0.29\textwidth}
    \centering
    \includegraphics[width=\textwidth]{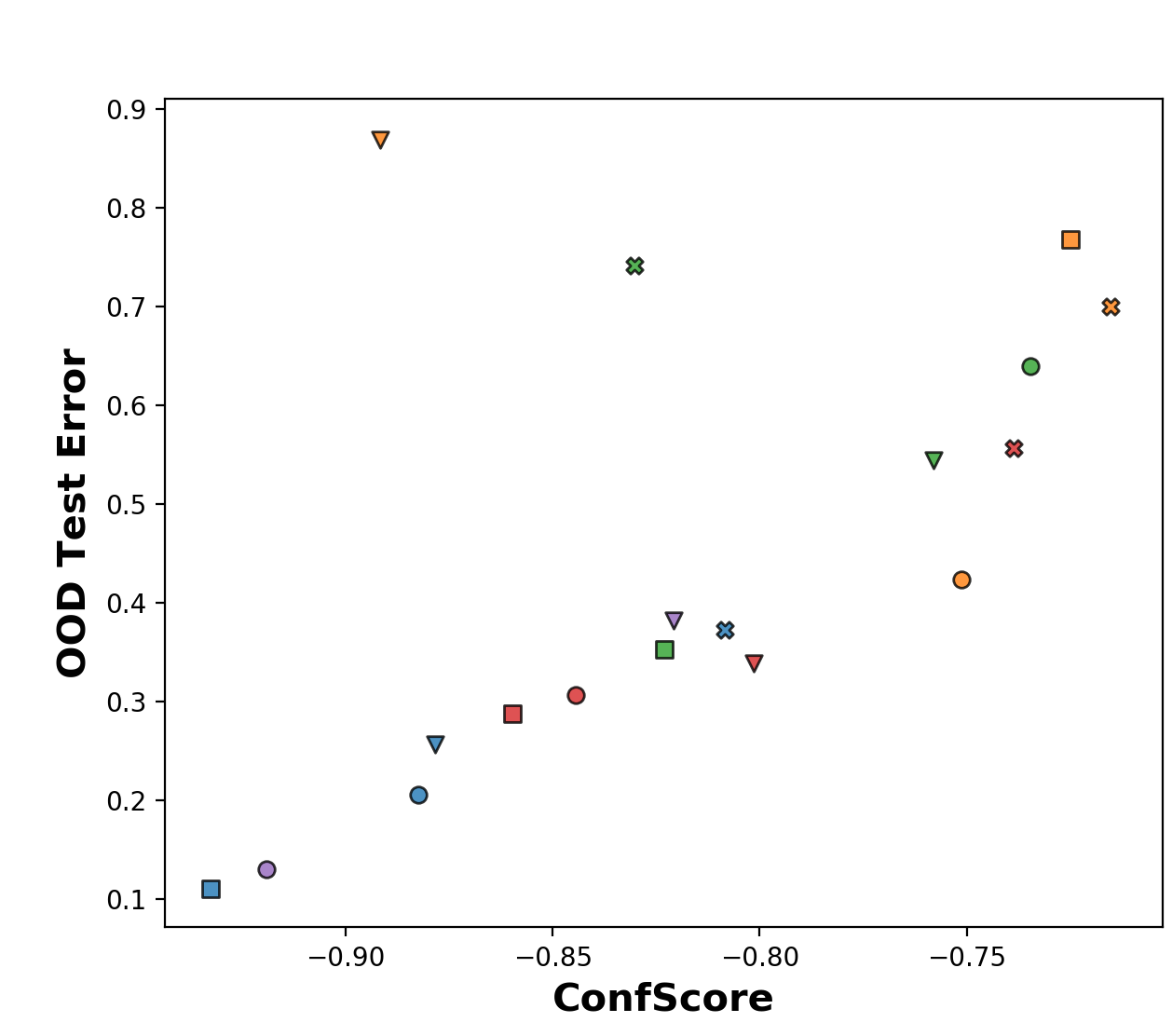}
    \end{minipage}
    \begin{minipage}{0.44\textwidth}
        \centering
    \includegraphics[width=\textwidth]{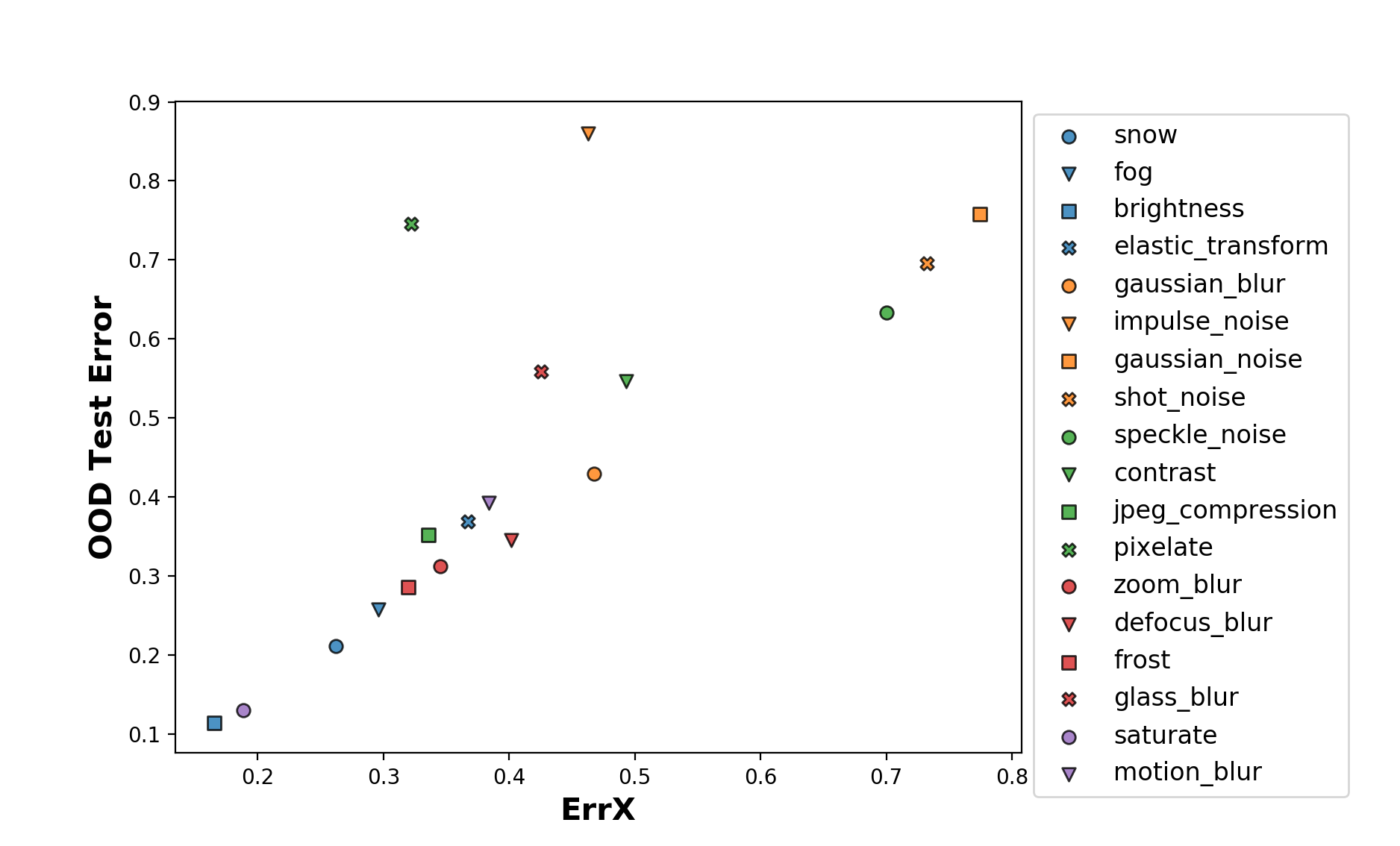}
    \end{minipage}
    \caption{Test error versus prediction on CIFAR10 with ResNet18 in the original common corruptions dataset. We plot the actual test errors on each corrupted dataset against predictions given by ProjNorm(left), ConfScore(middle), and ErrX(right).}
    \label{cc_original}
\end{figure}

\begin{figure}[H]
\begin{minipage}{0.29\textwidth}
         \centering
    \includegraphics[width=\textwidth]{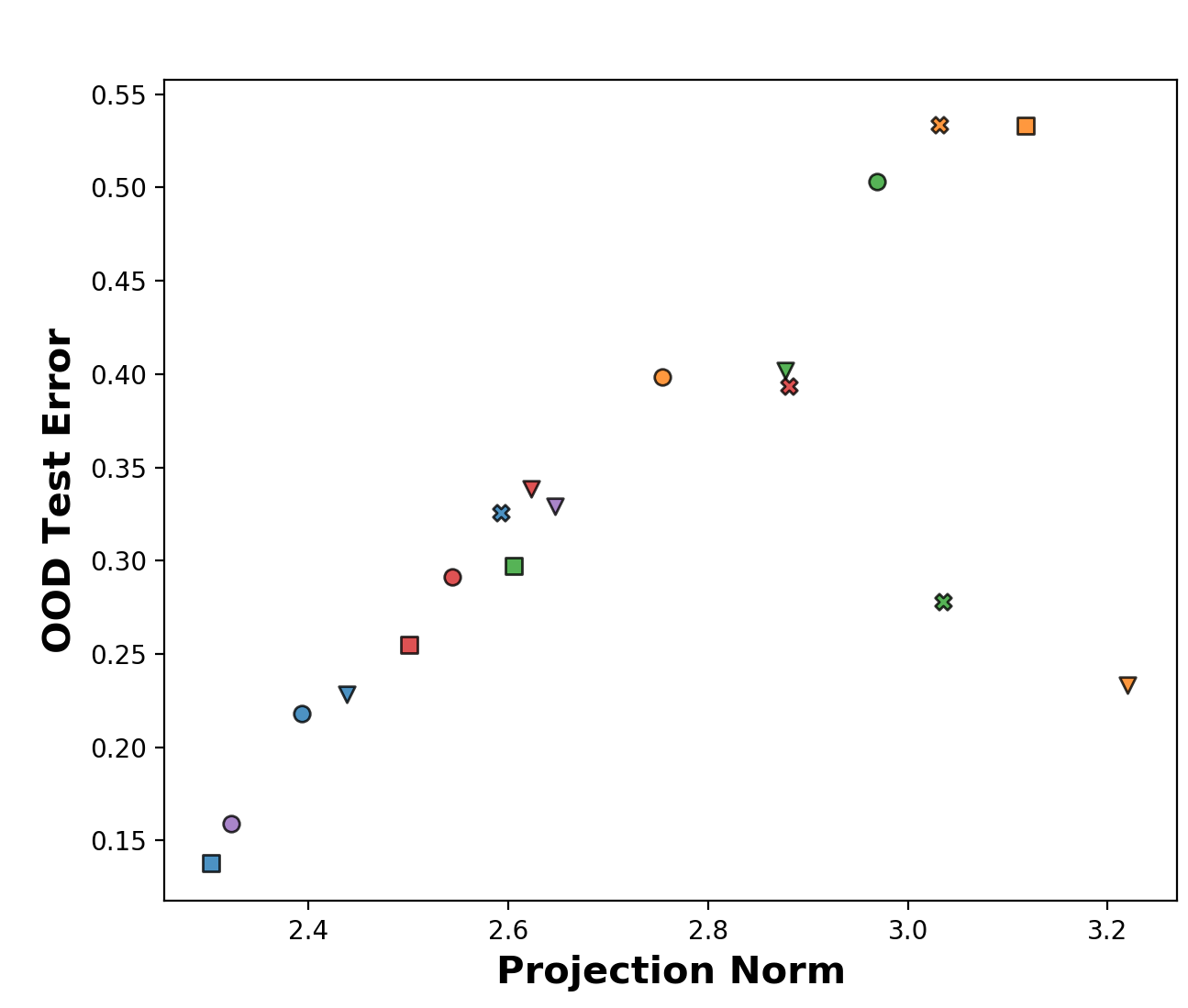}
    \end{minipage}
    \begin{minipage}{0.28\textwidth}
    \centering
    \includegraphics[width=\textwidth]{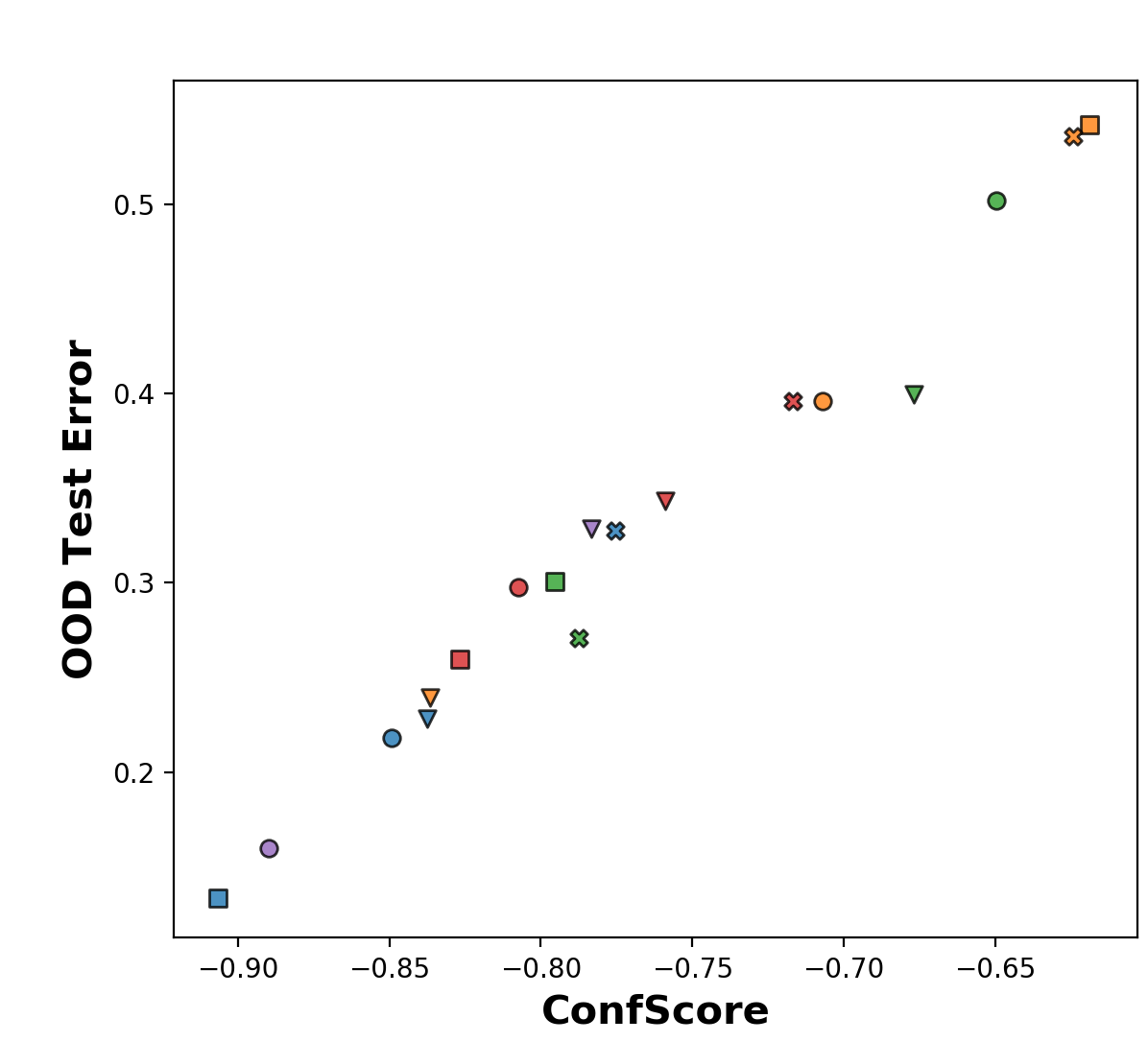}
    \end{minipage}
    \begin{minipage}{0.44\textwidth}
        \centering
    \includegraphics[width=\textwidth]{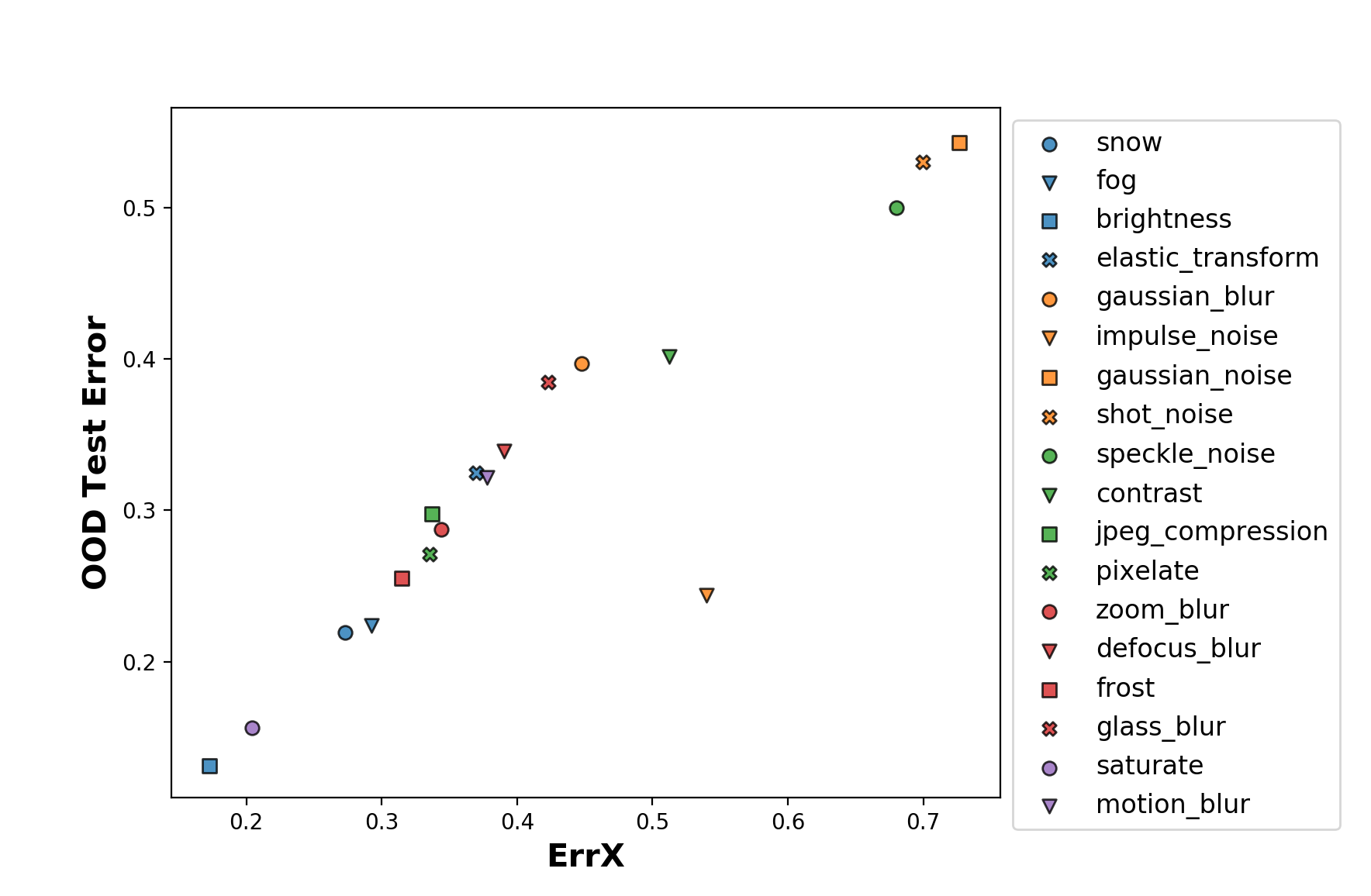}
    \end{minipage}
    \caption{Test error versus prediction on CIFAR10 with ResNet18 in the adapted common corruptions dataset with covariate shift. We plot the actual test errors on each corrupted dataset against predictions given by ProjNorm(left), ConfScore(middle), and ErrX(right).}
    \label{cc_adapted}
\end{figure}

The comparisons between actual test error and predictions for both the original and adapted versions of common corruptions dataset are presented in Figures \ref{cc_original} and \ref{cc_adapted}, respectively. It can be seens that the method of ErrX has similar performance in terms of correlation with actual test errors as compared to the other two methods that specialize in classification tasks. While the method of ErrX is computationally slower when calculating estimates using parametric bootstrap, it saves computation time by avoiding the task of calibration. Among the three predictions of test error, only ErrX is a direct estimate while the other two need calibration to match ProjNorm/Confscore to final error predictions, where calibration parameters may differ depending on training data, neural network architecture, etc.

\section{Discussion}
We propose an alternative method to estimate test error with observed covariate shift based on the target of average test errors over different potential training outcomes. Under the assumption of no conditional distribution shift, we provide two ways of estimation $\errX.dir$ and $\errX.dec$, based on parametric bootstrap. Unlike previous methods based on importance weighting, our method is not limited by the amount of overlap or shift between training and test covariates. Empirically, both $\errX.dir$ and $\errX.dec$ demonstrate consistent advantage over cross-validation under covariate shift across many modeling assumptions for both simulation data and real data example. While $\errX.dec$ performs slightly better than $\errX.dir$ in simulation settings, the two estimators have similar results in real data examples. 

However, the performance of $\errX.dir$ and $\errX.dec$ is heavily dependent on obtaining a good parametric model. In the presence of regularization, debiasing corrections are needed for better prediction accuracy. Besides, computation cost may also be a concern, especially for models that are hard to train since refitting in parametric bootstrap is involved. The assumption of no conditional distribution shift is also sometimes difficult to guarantee in practice. In the real data example of crime rate prediction, the estimators $\errX.dir$ and $\errX.dec$ do not fully recover the actual test errors, but consistently show improvements over cross-validation despite underlying model misspecification and potential conditional distribution shift. The advantage of $\errX.dir$ and $\errX.dec$ over CV grows when the amount of covariate shift is larger.

% \clearpage
\singlespacing
\bibliographystyle{alpha}
\bibliography{ref}

\begin{appendix}
\section{Proof of results}
\subsection{Proof of Proposition \ref{OLS_formula}}
\vspace{5 mm}\noindent{\bf Proposition 1}. {\it 
For linear models with the OLS fitting algorithm and squared error loss, assume in addition that the test data covariates are standardized so that $\mathbb{E}_{Q}[X] = 0$ and $\mbox{Var}_{Q}[X] = \Sigma$ of full rank. Then 
\begin{equation*}
    \errX^Q = \sigma^2 + \frac{\sigma^2}{n} \mbox{tr}\left(\hat{\Sigma}^{-1} \Sigma\right) = \errin(\bX) + \frac{\sigma^2}{n} \left(\mbox{tr}\left(\hat{\Sigma}^{-1}\Sigma \right) - p\right),
\end{equation*}
where $\hat{\Sigma} = \frac{1}{n} \sum_{i=1}^n x_i x_i^T$ is empirical covariance for training data.}

\begin{proof}
The proof is almost identical to that of Proposition 2 in \cite{bates2021cross} except that the target $\errX^Q$ depends on the test data distribution $Q$, which may be different from that of training data. Notice that by the decomposition formula,
\begin{align*}
    \errX^Q &= \errin(\bX) +  \E_{y \sim P_{Y|X= \bX}} \E_{y_0 \sim P_{Y|X=x_0}} \left[\E_{x_0 \sim \delta_{\bX_{test}}}[(y_0- \hat{f}(x_0,\hat{\theta}(\bX,y)))^2]
 - \E_{x_0 \sim \delta_{\bX}} [(y_0- \hat{f}(x_0,\hat{\theta}(\bX,y)))^2] \right]\\
 &= \errin(\bX) + \mathbb{E}_{y \sim P_{Y|X=\bX}}\left[\left(\theta - \hat{\theta}(\bX,y)\right)^T \Sigma \left(\theta - \hat{\theta}(\bX,y)\right) - \left(\theta - \hat{\theta}(\bX,y)\right)^T \hat{\Sigma} \left(\theta - \hat{\theta}(\bX,y)\right)\right]\\
 &= \errin(\bX) + \sigma^2 \mbox{tr}\left[ (\Sigma - \hat{\Sigma}) (\bX^T\bX)^{-1}\right] \\
 &= \errin(\bX) + \frac{\sigma^2}{n}\left(\mbox{tr}(\hat{\Sigma}^{-1}\Sigma) -p\right).
\end{align*}
Similarly, we get by direct expression of $\errX^Q$ that 
\begin{align*}
    \errX^Q &= \E_{y \sim P_{Y|X= \bX}} \left[\E_{(x_0,y_0) \sim Q}[(y_0- \hat{f}(x_0,\hat{\theta}(\bX,y)))^2]\right]\\
    &= \sigma^2 + \E_{y \sim P_{Y|X= \bX}}\left[\left(\theta - \hat{\theta}(\bX,y)\right)^T \Sigma \left(\theta - \hat{\theta}(\bX,y)\right)\right]\\
    &= \sigma^2 + \sigma^2 \mbox{tr}\left[\Sigma (\bX^T\bX)^{-1}\right]\\
    &= \sigma^2 + \frac{\sigma^2}{n}\mbox{tr}(\hat{\Sigma}^{-1}\Sigma).
\end{align*}
\end{proof}

\subsection{Proof of Proposition \ref{OLS-unbiased}}

\vspace{5 mm}\noindent{\bf Proposition 2}. {\it 
For linear models with the OLS fitting algorithm and squared error loss, assume in addition that the test data covariates are standardized so that $\mathbb{E}_{Q}[X] = 0$ and $\mbox{Var}_{Q}[X] = \Sigma$ of full rank. Then the estimators $\errX.dir$ and $\errX.dec$ in the algorithms \ref{errx1} and \ref{errx2} are unbiased for the target $\errX^Q$.}
\begin{proof}
Since we are given an unbiased estimator of in-sample error $\widehat{\errin}(\bX)$, it suffices to show that, $\frac{1}{B}\sum_{b=1}^B \widehat{\errX}^{(b)}$ is an unbiased estimator for $\errX^Q$, and that $\frac{1}{B}\sum_{b=1}^B \widehat{\errin}^{(b)}(\bX)$ is an unbiased estimator for $\errin(\bX)$, respectively. In other words, we want to show that for any $\eta \sim \mathcal{N}(0, I_n \sigma^2)$,

$$
\E_{y^* = \bX \hat{\theta}(\bX,\by) + \eta}\left[\E_{x_0 \sim Q_X}\left(x_0 \hat{\theta}(\bX,\by) - \hat{f}(x_0, \hat{\theta}(\bX, y^*))\right)^2\right] = \frac{\sigma^2}{n}\mbox{tr}(\hat{\Sigma}^{-1} \Sigma), 
$$
and
$$
\E_{y^* = \bX \hat{\theta}(\bX,\by) + \eta}\left[\E_{x_0 \sim \delta_\bX}\left(x_0 \hat{\theta}(\bX, \by) - \hat{f}(x_0, \hat{\theta}(\bX, y^*))\right)^2\right] = \frac{\sigma^2}{n}p.
$$

We will begin by proving the first equality. Let $M = \bX(\bX^T \bX)^{-1} \bX^T$ be the hat matrix. Since $x_0 \sim Q_X$, $\E(x_0) = 0$ and $\var(x_0) = \Sigma$, we have that,
\begin{align*}
    & \quad \E_{y^* = \bX \hat{\theta}(\bX,\by) + \eta}\left[\E_{x_0 \sim Q_X}\left(x_0 \hat{\theta}(\bX,\by) - \hat{f}(x_0, \hat{\theta}(\bX, y^*))\right)^2\right] \\
    &= \E_{y^* = \bX \hat{\theta}(\bX, \by) + \eta} \left[\E_{x_0 \sim Q_X} \left[\left(x_0 (\bX^T\bX)^{-1} \bX^T (\by - y^*)\right)^2\right] \right] \\
    &= \E_\eta \left[\E_{x_0 \sim Q_X} \left[\left(x_0 (\bX^T\bX)^{-1} \bX^T ((I-M)\by -\eta)\right)^2\right]\right] \\
    &= \E_\eta \left[\mbox{Var}_{x_0 \sim Q_X} \left[x_0 (\bX^T\bX)^{-1} \bX^T ((I-M)\by -\eta)\right]\right] \\
    &= \E_{\eta}\left[((I-M)\by-\eta)^T \bX(\bX^T\bX)^{-1} \Sigma (\bX^T\bX)^{-1}\bX^T((I-M)\by - \eta)\right]\\
    &= \E_\eta \left[\mbox{tr}\left[\bX (\bX^T\bX)^{-1} \Sigma (\bX^T\bX)^{-1}\bX^T\eta\eta^T\right]\right]\\
    &= \sigma^2 \mbox{tr}\left[\Sigma(\bX^T\bX)^{-1}\right]\\
    &=\frac{\sigma^2}{n} \mbox{tr}(\hat{\Sigma}^{-1}\Sigma),
\end{align*}
where the third last step is due to 
\begin{align*}
& ((I-M) \by)^T  \bX \left(\bX^T \bX \right)^{-1} \Sigma\left(\bX^T \bX \right)^{-1}\bX^T((I-M) \by )  = 0. 
\end{align*}

Similarly, the second equality as follows. 
\begin{align*}
    \E_{y^* = \bX \hat{\theta}(\bX,\by) + \eta}\left[\E_{x_0 \sim \delta_\bX}\left(x_0 \hat{\theta}(\bX, \by) - \hat{f}(x_0, \hat{\theta}(\bX, y^*))\right)^2\right]  &= \sigma^2 \mbox{tr}\left[\hat{\Sigma} (\bX^T\bX)^{-1}\right]
    = \frac{\sigma^2}{n}p.
\end{align*}

\end{proof}

\section{Algorithm for estimating in-sample error}
\label{in-sample}
In Section \ref{method}, we mentioned that \cite{ye1998measuring} and \cite{efron2004estimation} give a general form of covariance penalty identity for in-sample error under square error loss,
\begin{equation}
\label{cp_covariance}
\errin(\bX) = \E\left[\frac{1}{n}\sum_{i=1}^n \left(y_i - \hat{f}(x_i, \hat{\theta}(\bX, \by))\right)^2\mid \bX\right] + \frac{2}{n} \sum_{i=1}^n \mbox{Cov} \left(y_i, \hat{f}(x_i, \hat{\theta}(\bX, \by))\mid \bX\right).
\end{equation}
This allows us to derive a general method for estimating in-sample error summarized in Algorithm \ref{algo-cov}. 
\begin{center}
\begin{minipage}{\linewidth}
\begin{algorithm}[H]
\caption{\em  Estimate of in-sample error under squared error}
\label{algo-cov}
\hspace*{\algorithmicindent} \textbf{Input}: training data $(\bX, \by)$, number of bootstrap samples $B$, fitting algorithm $\hat{\theta}(\cdot)$, parametric model $P^\theta_{Y|X}$
\begin{algorithmic}[1]
\State Fit a model on training data to obtain $\hat{\theta}(\bX, \by)$.
\For{each $b\in\{1,\ldots, B\}$}
\State Generate vectors of outcomes for training data with parameter $\hat{\theta}(\bX,\by)$
\begin{align*}
    & \by^{(b)} \sim P_{Y|X = \bX}^{\hat{\theta}(\bX,\by)}. 
\end{align*}
 
\State Refit a model on bootstrap sample $\bX$, $\by^{(b)}$ to obtain $\hat{\theta}^{(b)} = \hat{\theta}(\bX, \by^{(b)})$. 
\EndFor
\For{each i = 1,\ldots, n}
\State Compute sample averages $\bar{Y}_i = \frac{1}{B}\sum_{b=1}^B \by^{(b)}_i$ and $\bar{f}_i = \frac{1}{B}\sum_{b=1}^B \hat{f}(x_i, \hat{\theta}^{(b)})$.
\State Compute $$\widehat{\mbox{Cov}}_i = \frac{1}{B}\sum_{b=1}^B \left(\by_i^{(b)} - \bar{Y}_i \right)\left(\hat{f}(x_i, \hat{\theta}^{(b)}) - \bar{f}_i\right)$$
\EndFor
\State Compute
$$
\widehat{\errin}(\bX) = \frac{1}{n}\sum_{i=1}^n \left(\by_i - \hat{f}(\bX_i, \hat{\theta}(\bX,\by))\right)^2 + \frac{2}{n}\sum_{i=1}^n \widehat{\mbox{Cov}}_i
$$
\end{algorithmic}
\hspace*{\algorithmicindent} \textbf{Output}: $\widehat{\errin}(\bX)$ 
\end{algorithm}
\end{minipage}
\end{center}

The covariance penalty based method of estimating in-sample can be generalized to a wider class of loss functions beyond squared error \cite{efron2004estimation}.
In the case of logistic regression with counting error, error function $q(u) = \mbox{min}(u, 1-u)$. We have the following identity,
\begin{equation*}
    \mathbb{E}[\errin]= \mathbb{E}\left[\frac{1}{n} \sum_{i=1}^n l(\by_i, \hat{f}(\bX_i, \hat{\theta}(\bX, \by)))\right] + 2\mbox{Cov}(\by_i, \lambda_i),
\end{equation*}
where $l$ is counting error function, $\lambda_i = -\frac{\partial{q}}{\partial{u}}(\hat{f}(\bX_i, \hat{\theta}(\bX, \by)))/2$. That is, $\lambda_i = -1/2$ if $\hat{f}(\bX_i, \hat{\theta}(\bX, \by)) = 0$ and $\lambda_i = 1/2$ otherwise. We shift all $\lambda_i$ by $1/2$ to get $\hat{\lambda}_i = \lambda_i + 1/2 = \hat{f}(\bX_i, \hat{\theta}(\bX, \by))$ without changing the covariance penalty term. Therefore, to estimate in-sample error for logistic regression with $\l$ being counting error loss, we only need to replace the last step in algorithm \ref{algo-cov} with
$$
\widehat{\errin}(\bX) = \frac{1}{n}\sum_{i=1}^n l\left(\by_i, \hat{f}(\bX_i, \hat{\theta}(\bX,\by))\right) + \frac{2}{n}\sum_{i=1}^n \widehat{\mbox{Cov}}_i. 
$$

\newpage
\section{Additional simulation results}
\label{table-simulation-others}

\begin{table}[!h]
\centering
\begin{tabular}{lccccc} 
\toprule
& \multirow{2}{*}{OLS}  & \multicolumn{2}{c}{Linear (Lasso)} & \multicolumn{2}{c}{Logistic (Lasso)} \\ \cmidrule(lr){3-4}\cmidrule(lr){5-6}
& & $p=10$  & $p=50$ & $p=10$   & $p=50$   \\\midrule
CV  & 0.713 & 0.352 & 0.323 & 0.257 & 0.423   \\ \midrule
$\errX.dir$  & 0.322 & 0.254 & 0.162 & 0.207 & 0.372\\ \midrule
$\errX.dec$  & 0.314 & 0.246 & 0.151 & 0.178 & 0.333  \\ \bottomrule
\end{tabular}
 \caption{\em  Comparison of mean squared difference between error estimates and actual test error for above simulation settings with covariate shift. The values in the table are standardized by the mean squared test error. Smaller absolute values are better. Multiplicative correction is used for linear regression with Lasso penalty, and relaxed Lasso correction is used for logistic regression.}
 \label{table-simulation-shift-squared}
\end{table}

\begin{table}[!h]
\centering
\begin{tabular}{lccccc} 
\toprule
& \multirow{2}{*}{OLS}  & \multicolumn{2}{c}{Linear (Lasso)} & \multicolumn{2}{c}{Logistic (Lasso)} \\ \cmidrule(lr){3-4}\cmidrule(lr){5-6}
& & $p=10$  & $p=50$ & $p=10$   & $p=50$   \\\midrule
CV  & 0.766 & 0.457 & 0.481 &  0.453 & 0.549  \\ \midrule
$\errX.dir$  & 0.466 & 0.402 & 0.340 &  0.329 & 0.485\\ \midrule
$\errX.dec$  & 0.465 & 0.391 & 0.328 &  0.329 & 0.480 \\ \bottomrule
\end{tabular}
 \caption{\em Comparison of mean absolute difference between error estimates and actual test error for above simulation settings with covariate shift. The values in the table are standardized by the mean test error. Smaller absolute values are better. Multiplicative correction is used for linear regression with Lasso penalty, and relaxed Lasso correction is used for logistic regression.}
 \label{table-simulation-shift-absolute}
\end{table}

\end{appendix}

\end{document}